\documentclass[12pt,twoside]{article}   %For printing on two sides of page
\usepackage[super,sort,comma]{natbib}
\usepackage{fancyhdr}		%Gives headers and footers defined below
\usepackage[section]{placeins}   %
\usepackage{graphicx}

\makeatletter \renewcommand\@biblabel[1]{$^{#1}$} \makeatother
\newcommand{\cen}[1]{\begin{center} #1 \end{center}}

\usepackage{hyperref}
\hypersetup{ colorlinks,
    citecolor=blue,
    filecolor=blue,
    linkcolor=blue,
    urlcolor=blue
}

\usepackage{xcolor}
\definecolor{gray}{rgb}{0.6,0.6,0.6}
\definecolor{red}{rgb}{0.85,0,0}
\definecolor{green}{rgb}{0,0.85,0}
\definecolor{blue}{rgb}{0,0,0.85}
\definecolor{beige}{rgb}{0.92,0.87,0.78}
\usepackage[all]{hypcap}    %causes link to figures to go to figure, not caption
\usepackage{caption}
\expandafter\let\csname equation*\endcsname\relax
\expandafter\let\csname endequation*\endcsname\relax
\usepackage{amsmath}
\usepackage{amssymb}
\usepackage{algorithm, algorithmicx}
\usepackage[noend]{algpseudocode}
\usepackage{multirow}
\usepackage{colortbl}
\usepackage{makecell}

\newcommand{\argmax}{\mathop{\mathrm{argmax}}}
\algrenewcommand\algorithmicrequire{\textbf{Input:}}  
\algrenewcommand\algorithmicensure{\textbf{Output:}}

\begin{document}

\cen{\sf {\Large {\bfseries High-dimensional Automated
      Radiation Therapy Treatment Planning via Bayesian Optimization \footnote{This is the pre-peer reviewed version of the following article: [Wang, Q., Wang, R., Liu, J., Jiang, F., Yue, H., Du, Y., \& Wu, H. (2023). High-dimensional automated radiation therapy treatment planning via bayesian optimization. Medical physics, 10.1002/mp.16289. Advance online publication. https://doi.org/10.1002/mp.16289], which has been published in final form at [DOI: 10.1002/mp.16289].} } \\  
    \vspace*{10mm}
    Qingying Wang$^{1,2,}$\footnotetext[2]{equal contribution} \footnotemark[2], Ruoxi
    Wang$^{1,}$ \footnotemark[2], Jiacheng Liu$^{1,2}$, Fan Jiang$^{1}$, Haizhen Yue$^{1}$, Yi Du$^{1,2,}$, Hao Wu$^{1,2,}$\footnote[3]{corresponding author}} \\
    $^1$ Key Laboratory of Carcinogenesis and Translational Research
    (Ministry of Education/Beijing), Department of Radiation Oncology, Beijing
    Cancer Hospital \& Institute, Beijing, China

    $^2$ Institute of Medical Technology, Peking University Health Science Center
  \vspace{5mm}\\
  Version typeset \today\\
}

\pagenumbering{roman}
\setcounter{page}{1}
\pagestyle{plain}
Author to whom correspondence should be addressed. email: hao.wu@bjcancer.org\\
% note, probably best not to use a student's e-mail as it won't be valid for
% very long.

\begin{abstract}
  \noindent
  {\bf Purpose:} Radiation therapy treatment planning can be viewed as
  an iterative hyperparameter tuning process to balance conflicting clinical
  goals. In this work, we investigated the performance
  of modern Bayesian Optimization (BO) methods on automated treatment planning problems
  in high-dimensional settings.\\
  {\bf Methods:} 20 locally advanced rectal
  cancer patients treated with intensity-modulated radiation therapy (IMRT) were
  retrospectively selected as test cases. The adjustable planning parameters
  included both dose objectives and their corresponding weights.
  We implemented an automated treatment planning framework and tested
  the performance of two BO methods on the treatment planning task: one
  standard BO method (GPEI) and one BO method dedicated to high-dimensional
  problems (SAAS-BO). A random tuning method was also included as the baseline. The three
  automated methods' plan quality and planning efficiency were compared with the
  clinical plans regarding target coverage and organs at risk (OAR) sparing.
  The predictive models in both BO methods were compared to analyze the
  different search patterns of the two BO methods.\\
  {\bf Results:} For the
  target structures, the SAAS-BO plans achieved comparable hot spot control ($p=0.43$) and
  homogeneity ($p=0.96$) with the clinical plans, significantly better than
  the GPEI and random plans ($p\textless0.05$). Both SAAS-BO and GPEI plans significantly
  outperformed the clinical plans in conformity and dose spillage ($p\textless0.05$).
  Compared with clinical plans, the treatment plans generated by the three
  automated methods all made reductions
  in evaluated dosimetric indices for the femoral head and the bladder.
  The analysis of the underlying predictive models has shown that both BO
  procedures have identified similar important planning parameters. \\
  {\bf Conclusions:} This work implemented a BO-based hyperparameter tuning
  framework for automated treatment planning. Both tested BO methods were able
  to produce high-quality treatment plans and reduce the workload of treatment planners.
  The model analysis also confirmed the intrinsic low dimensionality of the tested
  treatment planning problems. \\
\end{abstract}

\newpage     %may or may not be needed

\tableofcontents

\newpage

\setlength{\baselineskip}{0.7cm}      %double spacing		

\pagenumbering{arabic}
\setcounter{page}{1}
\pagestyle{fancy}
\section{Introduction}
The inverse planning in radiation therapy is achieved by minimizing a mathematical
objective function with regard to the beam arrangements\cite{bortfeld2006imrt}.
Traditionally, the treatment planner
sets the planning parameters formulating the objective function, in order to
satisfy multiple clinical goals. To find the optimal plan for an individual
patient, the planner has to modify multiple planning parameters in a
trial-and-error process to balance multiple conflicting clinical goals. When the number of adjustable parameters is
large, performing the tuning process mentioned above is time-consuming. In addition,
the planner's experience and planning time constraints contribute to
inconsistent plan qualities in the manual planning process. Previous studies have
shown that automated treatment 
planning may mitigate the shortcomings of manual planning approaches\cite{wu2003treatment}.
In the literature, three major directions in automated treatment planning can be listed:
\begin{itemize}
\item 
The \textbf{knowledge-based planning} method leverages historical planning data to
predict the dose-volume histograms (DVH), dose distributions, or optimization
objectives. The predicted DVH/dose distributions of a new patient
can serve as a warm start for the following optimization\cite{fredriksson2012automated,
  mcintosh2017fully, fan2019automatic}.
Another line of research aims to retrieve the planning parameters given the final
treatment plan via the inverse optimization (IO) method\cite{chan2014generalized,babier2018inverse, boutilier2015models}.
\item
The \textbf{multi-criteria optimization} (MCO) aims to handle conflicting
objectives\cite{craft2006approximating, bokrantz2013distributed} by searching for 
Pareto front for multiple objectives, instead of considering a composite scalar
objective. The MCO algorithms can be separated based on the final plan decision strategy\cite{bokrantz2013multicriteria}
into \textit{a posteriori} MCO\cite{hong2008multicriteria} and \textit{a
  priori} MCO\cite{breedveld2007novel}. In either approach, sensitivity
analysis of the objective weights or the constraint parameters is performed to
quantify the parameter influence on the objective function\cite{alber2002tools, sobotta2008tools}. However, this
requires the knowledge of the specific form of the objective function for
efficient derivative evaluation, which is not available in commercial treatment
planning systems (TPS).
\item
The \textbf{automated iterative planning} formulates the automated planning as an iterative
hyperparameter tuning process. 
A scoring function evaluating the current plan quality is defined, and the
planning parameters for the next optimization cycle are
adjusted\cite{xing1999optimization} according to previous optimization results.
Numerous works have embraced this idea due to its simplicity and minimum
interfacing requirement\cite{wang2021tree, zhang2011methodology, huang2022meta, shen2020operating}.
\end{itemize}
% position of the work compared with Maass.
Following the line of research in automated iterative planning, our work in this
paper involves applying Bayesian Optimization (BO) in treatment
planning. BO is a model-based hyperparameter optimization method used in
a wide range of areas, including automatic machine learning
\cite{turner2021bayesian}, robotics \cite{yuan2019bayesian}, sensor networks
\cite{sambito2021strategies}, and engineering design \cite{lam2018advances}. In
radiation treatment planning, BO has been applied to solve beam angle
optimization\cite{taasti2020automating} and hyperparameter
tuning problems\cite{landers2018fully, maass2022hyperparameter}.

Although the performance of BO in these missions demonstrates some potential,
the problem's dimensionality in the aforementioned works has been kept relatively
moderate ($\leq 20$). In clinical practice, the number of planning parameters
can mount up to 50, posing issues for the iterative planning
methods. Lu et al. detected sensitive parameters
according to the correlation measures
between each planning parameter and the composite plan quality\cite{lu2007reduced}. 
However, they identified the sensitive parameters based on a pre-calculated
optimization dataset, limiting the application scope of the methods.

% points to improve the performance of BayOpt in treatment planning.
The work most similar to ours is Maass et al.\cite{maass2022hyperparameter},
where the authors performed BO on the planning parameters for lung
cancer SBRT planning.
In this work, we further investigated the BO applications on automated treatment
planning in high-dimensional settings. We implemented a hyperparameter optimization
framework integrated with an Eclipse TPS (Varian Medical Systems, Inc., Palo Alto, CA). Based on this framework, we compared the performance of a recently proposed BO method for high-dimensional problems\cite{eriksson2021high}
with a standard BO method in the context of rectal cancer intensity-modulated
radiation therapy (IMRT) planning. Little expert knowledge was incorporated in
the planning problem design, to fully demonstrate the potential of the BO
applications in automated treatment planning. 
We demonstrate that both BO methods achieved
comparable or better plan quality in organs at risk (OAR) sparing and target
coverage than manual planning.
In addition, an analysis was performed on both BO method's predictive models.
The analysis demonstrated that both BO methods identified similar sensitive
planning parameters on-the-fly, without knowledge of prior optimized plans.
Based on the the prediction model analysis, an explanation was provided on the
different search behaviors of the two BO approaches. 

\section{Methods}
\label{sec:2}

\begin{figure}[ht]
  \centering
  \captionsetup{justification=justified}
  \includegraphics[width=\textwidth]{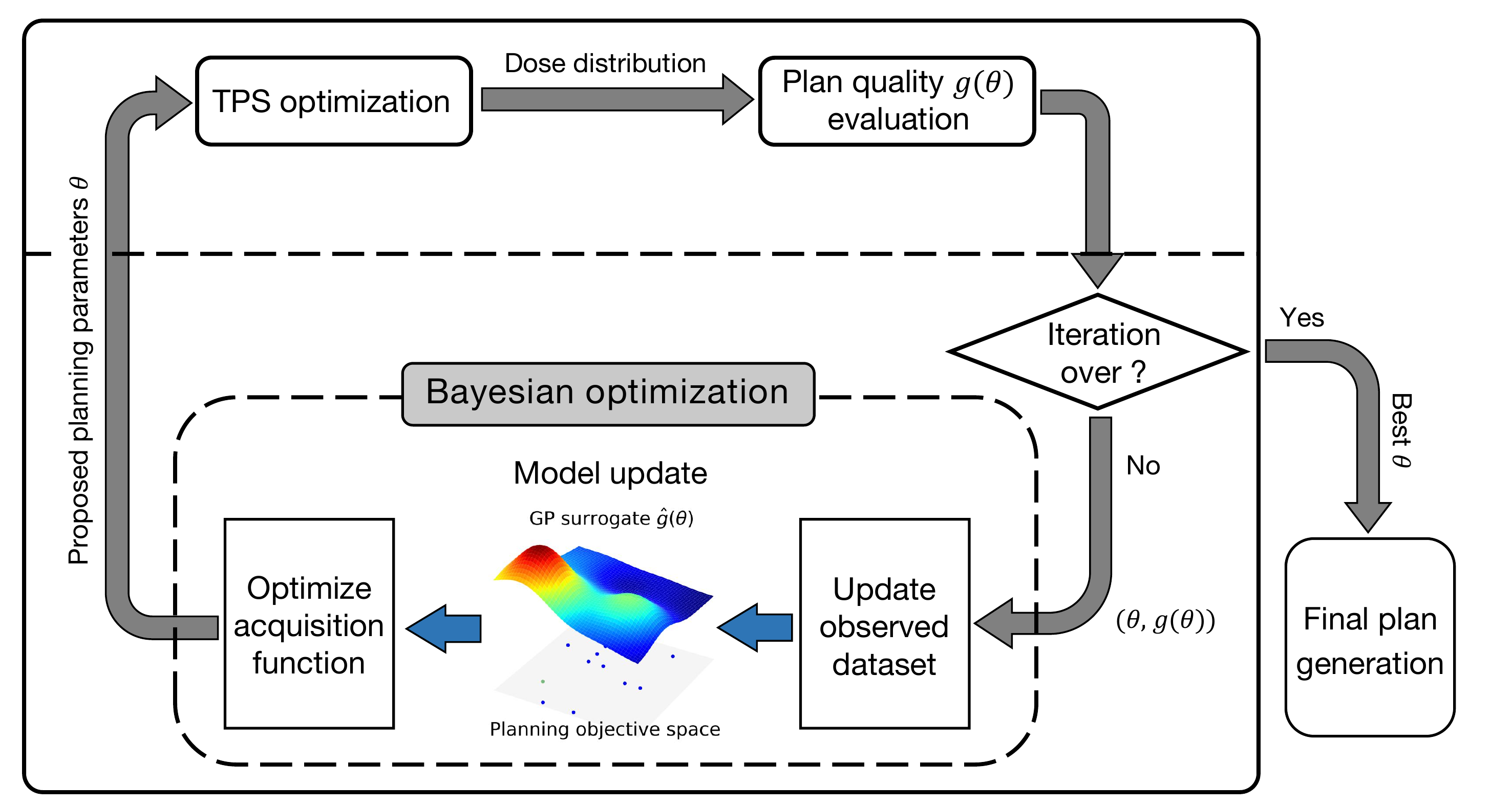}
  \caption{The framework of BO approach applied to automatic treatment planning.}
  \label{fig:1}
\end{figure}

The general scheme of the automated treatment planning
framework performed in this work is presented in figure \ref{fig:1}.
For clarity, we omitted the processes prior to the planning parameter setup in
the presented scheme (image registration, contouring, beam angle
optimization, etc.).
The main loop of this framework can be separated into two parts:
\begin{enumerate}
\item The first part involves emulating the manual planning process and the plan
  quality evaluation. A set of dose objectives and weights  $\theta = \{w_i,
  i=1, \ldots, s\}$ is sent to the TPS optimizer interface, specifying the fluence map
  optimization (FMO) problem to be solved. From the perspective of the FMO
  problem, $\theta$ is viewed as the hyperparameter set.
  Given the specific problem, the TPS
  optimizer outputs the optimal fluence map and the corresponding dose
  distribution. The plan quality was quantified by a predefined score
  function based on the corresponding dose distribution. The whole process is viewed as a 
  black-box function $g(\theta)$, where the input is the planning parameters and the output
  is the corresponding plan quality score.
  At last, the newly observed $(\theta, g(\theta))$ are sent into the BO
  procedure.
\item The second part consists of employing BO to find the most promising
  $\theta$ for the next round of TPS optimization. Specifically, the
  BO procedure appends the newly observed data $(\theta, g(\theta))$ to the
  observed dataset, updates the surrogate model with the new data set, and
  proposes the next hyperparameter set for trial. 
\end{enumerate}

Aside from the main loop in the framework, an exit
control has been included conditioned on the iteration numbers.
The final treatment plan was generated
with the currently best observed $\theta$, when the iteration number exceeds the
predefined threshold. In the remainder of this section, we first detail the BO
application in the automated treatment planning context and explain the
methodology we adopted for high-dimensional BO in subsection \ref{sec:2.1} and
\ref{sec:2.2}. The specific form of the plan quality score function is defined
in subsection \ref{sec:2.3} and the following subsections describe the
implementation and the experimental setup.

\subsection{Bayesian Optimization}
\label{sec:2.1}

As shown in figure \ref{fig:1}, the proposed framework
employs BO to select the planning parameters for treatment planning. Because
evaluating $g(\theta)$ is expensive, we want to rely on a sample-efficient meta
optimization method, as is the BO. As a model-based optimization algorithm, BO
tunes the planning parameters in two main steps:
\begin{enumerate}
\item \textbf{Modelling}: A surrogate model $\hat{g}(\theta)$ is used to
  approximate $g(\theta)$, given an observed dataset $\mathcal{D} = \{\theta_i,
  g(\theta_i) \}_{i=1}^t$. The arguably most popular surrogate model in BO is
  the Gaussian Process (GP)\cite{williams2006gaussian}.
  GP can be thought of as a Gaussian distribution of
  functions, characterized by a mean function $\mu$, and a covariance
  function $k(\theta, \theta')$. Without loss of generality, $\mu$ is set to 0,
  and the covariance function serves to describe the similarity between the
  function values ($\hat{g}(\theta)$ and $\hat{g}(\theta')$) as a function of
  the distance measure between two hyperparameter sets $r(\theta, \theta')$.
  In this work, a GP regression model equipped with a Mat\'{e}rn-5/2
  covariance/kernel function was used to approximate the score function $g(\theta)$, expressed as
  \begin{eqnarray}
    \label{eq:1}
    \mathrm{distance}: &r(\theta, \theta')^2&= \sum \limits_{i=1}^s \left ((w_i
                                               - w_i') / l_i\right)^2,\nonumber\\
    \mathrm{covariance}: &k(\theta, \theta')&= \sigma_k^2(1 + \sqrt{5}r + \frac{5}{3}r^2 )\exp(-\sqrt{5}r),\nonumber\\
    \mathrm{function\ value}: & \hat{g}(\theta) &\sim \mathcal{GP}(0,
                                                         k(\theta, \theta')),
                                                         \nonumber \\
    \mathrm{observation}: &  g(\theta) &\sim
                                                \mathcal{N}(\hat{g}(\theta),
                                                \sigma^2),
  \end{eqnarray}
  where $\sigma^2$ denotes the observational noise, $\sigma_k^2$ the
  kernel variance. We considered the optimization environment noiseless,
  so $\sigma^2$ was set to $1\times 10^{-6}$ for regularization purposes.
  In the distance measure,  the length scale $l_i$ was
  introduced to control the relevance of the corresponding planning parameter
  $w_i$. Generally, larger $l_i$ indicates less relevance of $w_i$ because
  identical variation with larger length scales corresponds to smaller changes in
  the distance measure. The GP model's hyperparameter set $\psi=\{\sigma_k,
  l_1, \ldots, l_s\}$ was determined by maximizing the model's probability with
  regard to the observed dataset. 
  % TODO: discuss where to put this sentence
  Since $\l_i$ depends on the range of $w_i$, we normalized $\theta$ to the $[0,
  1]^s$ domain before learning the GP model to ensure robust model
  performance. Note that at the initial stage of BO, the observed dataset was
  populated by pseudo-random hyperparameter sets $\theta_{1:m}$ and their corresponding
  plan quality scores.
  
\item \textbf{Acquisition function}: When the GP regression model was learned, an acquisition
  function was used to find the next hyperparameter set for trial.
  While various acquisition functions have been proposed, we used in this work
  the expected improvement (EI) due to its computational efficiency and good empirical
  performance\cite{jones1998efficient}. The EI is defined as
  %\begin{linenomath}
  \begin{equation}
    \label{eq:2}
    \mathrm{EI}(\theta) = \mathbb{E}[\max(0, g(\theta) - g(\theta^+))],
  \end{equation}
  %\end{linenomath}
  where $\mathbb{E}$ denotes the expected value, and $g(\theta^+)$ denotes
  the currently best observed value. Note that the $g(\theta)$ prediction
  incorporates both the mean and the variance, due to the probabilistic nature
  of the GP regression model. Therefore the EI inherently balance between
  exploration and exploitation. The hyperparameter set with the highest EI is
  then selected for the next trial. A summary of the described BO
  procedure is given in Algorithm \ref{algo:1}.
\end{enumerate}

\begin{algorithm}
  \caption{BO procedure for treatment planning}
  \label{algo:1}
  \begin{algorithmic}[1]
    \Require{score function $g(\cdot)$; initial evaluation budget $m$; total
      evaluation budget $T$.}
    \Ensure{final maximizer and maximum ($\theta_{\max}, g(\theta_{\max})$)}
    \State Sample initial hyperparameter set $\theta_{1:m}$.
    \State Query $g(\theta_i), i = 1, \ldots, m$ by passing $\theta_i$ to TPS for optimization.
    \State Let $\mathcal{D} = \{(\theta_i, g(\theta_i))\ \mathrm{for}\ i = 1, \ldots, m\}$.
    \For {$t = m+1, \ldots, T$}
    \State Fit GP with $\mathcal{D}$ to determine $\psi_t$.
    \State Optimize EI to obtain $\theta_t =
    \argmax_\theta{\mathrm{EI}(\theta \vert \psi_t)}$.
    \State Query $g(\theta_t)$.
    \State Append ($\theta_t, g(\theta_t)$) to $\mathcal{D}$.
    \State $\theta_{\max} = \argmax \{g(\theta_{i})\ \mathrm{for}\ i = 1,
      \ldots, t)\}$.
    \EndFor 
    \State \Return ($\theta_{\max}, g(\theta_{\max})$)
  \end{algorithmic}
\end{algorithm}

\subsection{High dimensional considerations}
\label{sec:2.2}

The dimension of the planning parameters in a realistic treatment planning problem varies
from 20 to 50, depending on the tumor sites and prescription complexities. The
high dimensionality poses a problem for many meta optimization
approaches. In the context of BO, we have applied a recently proposed BO
algorithm named ``Sparse Axis Aligned Subspace BO''(SAAS-BO)\cite{eriksson2021high} to solve the high-dimensional treatment planning problems. 
The core idea of SAAS-BO is to introduce a sparsity-inducing prior on the GP
length scales $\{l_i\}_{i=1}^s$, assuming that only a few parameters are important for
the composite plan quality. This assumption is in accordance with previous findings
that the effective dimension in treatment planning problems can be much smaller
than the number of adjustable parameters\cite{lu2007reduced}. 
We chose this algorithm due to its great empirical performance and the
model's explicability since the dimension reduction
was performed on the original axis of the planning parameters.
The interested reader is referred to the cited reference for a detailed
description of this method.
To evaluate the SAAS-BO's effectiveness in automated treatment planning, we
have included both a standard BO (GPEI)  and a random-sampling method (random)
as baseline methods. In the GPEI approach, a uniform gamma prior was placed on
all the length scales of the GP regression model. The EI was used as the
acquisition function, identical to SAAS-BO. In the random sampling method, a
Sobol\cite{bratley1988algorithm} sequence was used to generate the random
planning parameters. The full GP regression model specifications in the
SAAS-BO and GPEI methods were listed in the supplementary material for reference.

\subsection{Score Function}
\label{sec:2.3}
To evaluate the agreement with the clinical goals, the score function was
defined as a weighted sum of multiple plan quality metrics (PQM):
\begin{eqnarray}
  \label{eq:3}
  g(\theta) =& \left(\sum\limits_{i=1}^{n}\alpha_iF_i(m_i;\ \bar{m}_i) \right) / \sum\limits_{i=1}^n \alpha_i,\\
  F_i(m_i; \bar{m}_i) =& \min\limits_{j=1,2}(a_j (m_i - \bar{m}_i) / \bar{m}_i)\
             \mathrm{and}\ \alpha_i =
             2^{-p_i}, \nonumber
\end{eqnarray}
where $F_i(m_i; \bar{m}_i)$ denotes the $i$th scoring term, $m_i$ the
corresponding PQM derived from the dose distribution generated by the planning parameter
set $\theta$, and $\bar{m}_i$ the clinical goal to achieve.
Specifically, we used 2-segment piecewise linear functions to characterize
$F_i(m; \bar{m}_i)$, which serves to penalize the score severely when the clinical
goal is not reached, and to reward slightly for the further improvement once the clinical goal is achieved.
$a_j$ defines the slope of each segment in the piecewise linear function.
Similar to Huang et al.\cite{huang2022meta}, we grouped the clinical goals into several
tiers indexed by $p_i$, and set the weight $\alpha_i$ associated with the $i$th goal as
an inverse exponential function of $p_i$. We set the scoring terms related to the target
structures to the top tier, to emphasize their importance in the plan evaluation.
The terms related to dose sparing for OAR and healthy tissues were
set to lower tiers, respectively. The specific formats of PQM terms
used in this study included min/max dose, mean dose, dose-volume parameters
(e.g., $D_{2\%}$ for hot spot control), and dose spillage ($R_{50\%},
R_{90\%}$)\cite{Dong2013}.  For the rectal cancer test cases, The clinical goals
related to these terms were set according to the institutional requirements\cite{li2012preoperative}, as listed in table \ref{tab:1}.

\begin{table}[ht]
\begin{center}
\arrayrulecolor{black}
\caption{Clinical goals and relevant parameters in utilized PQM terms.}
\label{tab:1}
\begin{tabular}{cccccccc} 
\arrayrulecolor{black}\hline
 & Type & $a_1$ & $a_2$ & Structure & Parameter & $\bar{m}_i$ & $p_i$ \\ 
\arrayrulecolor{black}\hline
\multirow{5}{*}{Target} & \multirow{4}{*}{lower} & \multirow{4}{*}{0} & \multirow{4}{*}{100} & GTV & $D_{\min}$ (Gy)& 50.6 & 1 \\
 &  &  &  & CTV~ & $D_{\min}$ (Gy) & 41.8 & 1 \\
 &  &  &  & PGTV & $D_{98\%}$ (Gy)& 50.6 & 1 \\
 &  &  &  & PTV~ & $D_{98\%}$ (Gy)& 41.8 & 1 \\ 
 & upper & 0 & -100 & PGTV & $D_{2\%}$ (Gy)& 55.154 & 1 \\ 
\cline{2-8}
\multirow{5}{*}{OAR} & \multirow{5}{*}{upper} & \multirow{5}{*}{-10} & \multirow{5}{*}{-100} & \multirow{2}{*}{Femoral Head} & $V_{20\mathrm{Gy}}$ (\%) & 50 & 3 \\
  &  &  &  &  & $V_{30\mathrm{Gy}}$ (\%)& 20 & 3 \\
  \cline{6-8}
 &  &  &  & \multirow{3}{*}{Urinary Bladder} & $V_{25\mathrm{Gy}}$ (\%)& 50 & 2 \\
 &  &  &  &  & $V_{45\mathrm{Gy}}$ (\%)& 20 & 2 \\
 &  &  &  &  & $D_{\mathrm{mean}}$ (Gy)& 18 & 2 \\ 
\cline{2-8}
\multirow{2}{*}{Dose spillage} & \multirow{2}{*}{upper} & \multirow{2}{*}{-10} & \multirow{2}{*}{-100} & \multirow{2}{*}{PTV~} & $R_{50\%}$ & 4.5 & 2 \\
 &  &  &  &  & $R_{90\%}$ & 1.5 & 2 \\
\hline
\end{tabular}
\end{center}
\end{table}
\subsection{Implementation}
\label{sec:2.4}
The proposed auto planning framework was coupled with an Eclipse TPS v16.1,
where the Python Eclipse Scripting Application Programming Interface
(\texttt{PyESAPI}) was used for message passing between the BO procedure and the
TPS. Within the Eclipse TPS, the Photon Optimizer (PO) was used
for inverse planning, while the intermediate and final doses were calculated with the Acuros XB
algorithm (AXB)\cite{vassiliev2010validation}. The voxel sizes for PO and AXB
were set at 2.5 mm, identical to the clinical configuration. All three
automated treatment planning procedures (random, GPEI, SAAS-BO) have been implemented based on the open-source
libraries \texttt{ax/Botorch}\cite{bakshy2018ae, balandat2020botorch}.
We set the iteration budgets $m = 20$ and $T = 120$, respectively, to control the optimization time clinically
relevant.
% further process for the clinical plans comparison
To compare with the clinical plans, the intermediate dose was calculated at the
end of the hyperparameter tuning process, and the final optimization was
performed based on the accurately calculated dose without changing any planning parameters.
% Clinical considerations on the normalization
The treatment plans were normalized to satisfy the institutional prescription on
target volumes. When the prescription contains multiple requirements, the smallest
normalizing factor was chosen to fulfill all target volume prescriptions.
To ensure this study as reproducible as possible, we have made the implementation
publicly available\footnote{https://github.com/inamoto85/BOPlanner}.

\subsection{Clinical Experiment}
\label{sec:2.5}
% clinical setup layout
The implemented iterative planning approach was tested on rectal cancer
treatment plans. We obtained 20 preoperative IMRT treatment plans for locally-advanced rectal cancer patients in our
institution. All treatment plans were manually optimized and previously treated.
The plans were all designed with the
Eclipse TPS to deliver equispaced 7-field IMRT plans
with 10MV photon beam. The target prescription implements a concomitant boost
technique with 22 fractions for the clinical target volume
($\mathrm{CTV}$) and the gross tumor volume
($\mathrm{GTV}$), respectively\cite{li2012preoperative}. Additional
target regions of interest (ROI) were created by expanding the aforementioned
structures with a 5 mm margin, identified as the PTV and the PGTV.
% insert a table of target volume prescription here?
An auxiliary structure named ``IrradVolume'' was created by subtracting the
PGTV from the PTV for additional dose distribution control.
% target prescription details
The institutional prescription requires 95\% and 99.9\% coverage of the PTV and
the CTV with 41.8 Gy, and similarly 95\% and 99.9\% coverage of the PGTV and the
GTV with 50.6 Gy\cite{li2012preoperative}.
The OARs included the bladder, femoral head, external body, and avoidance, which is
a support structure to spare both the anterior and posterior abdomen, covering mainly the
small intestine and the bone marrow.
% dose objectives
The optimization objectives for the ROIs included max/min dose, mean dose, and
dose-volume parameters. Both the dose objective and the corresponding weight
were considered as adjustable parameters, formulating a 34-D problem. The adjustment
range of the weights was defined uniformly as $[100, 1000]$. However, the
target dose objective ranges were defined as $[100\%, 105\%]$ of the
respective dose prescriptions.
The upper dose limits for the OARs were defined according to RTOG-0822
\cite{hong2015nrg} and institutional experiences, while the lower dose limits for the OARs
were set uniformly at 1 Gy. In addition, constraints on the dose objectives were
introduced to avoid logically contradictory objectives, e.g.,
$D_{\min}^{\mathrm{GTV}} < D_{\max}^{\mathrm{Body}}$.
Table \ref{tab:2} details all the parameters, adjustment ranges, and constraints used for
the two BO approaches. Finally, the normal tissue objective (NTO) was set with a fixed weight of
500 to control dose fall-off.
\begin{table}[ht]
\begin{center}
\arrayrulecolor{black}
\caption{Optimization parameters, ranges and constraints of rectal cancer cases
  treatment planning. Abbreviations: IV - IrradVolume; FH - Femoral Head; UB:
  Urinary Bladder.}
\label{tab:2}
\begin{tabular}{ccccc}
\arrayrulecolor{black}\hline
Parameter & Type & Dose Objective (Gy) & Weight & Constraints\\ 
  \arrayrulecolor{black}\hline
  $D^{\mathrm{Body}}_{\max}$ & upper & [50.6,53.13] & \multirow{17}*{[100,1000]}&  \\
  $D^{\mathrm{CTV}}_{\min}$ & lower & [41.8,43.89] & & \\
  $D^{\mathrm{PTV}}_{\min}$ & lower & [41.8,43.89] & & $D^{\mathrm{PTV}}_{\min} \leq D^{\mathrm{CTV}}_{\min}$\\
  $D^{\mathrm{GTV}}_{\min}$ & lower & [50.6,53.13] & & $D^{\mathrm{GTV}}_{\min} \leq D^{\mathrm{Body}}_{\max}$\\
  $D^{\mathrm{PGTV}}_{\min}$ & lower & [50.6,53.13] & & $D^{\mathrm{PGTV}}_{\min} \leq D^{\mathrm{GTV}}_{\min}$  \\

   $D^{\mathrm{IV}}_{\max}$ & upper & [44.65,49.35] & & \\
  $D^{\mathrm{IV}}_{\min}$ & lower & [41.8,43.89] & & \\
  $D^{\mathrm{Avoidance}}_{\max}$ & upper & [1,45] & & \\
  $D^{\mathrm{Avoidance}}_{\mathrm{mean}}$ & upper & [1,20] & & $D^{\mathrm{Avoidance}}_{\mathrm{mean}} < D^{\mathrm{Avoidance}}_{\max}$\\ 
  $D^{\mathrm{FH}}_{\max}$ & upper & [1,50] & & \\
  $D^{\mathrm{FH}}_{25\%}$ & upper & [1,20] &  & $D^{\mathrm{FH}}_{25\%} < D^{\mathrm{FH}}_{\max}$\\
  $D^{\mathrm{FH}}_{40\%}$ & upper & [1,20] &  & $D^{\mathrm{FH}}_{40\%} < D^{\mathrm{FH}}_{25\%}$\\
  $D^{\mathrm{FH}}_{\mathrm{mean}}$ & upper & [1,15] &  & $D^{\mathrm{FH}}_{\mathrm{mean}} < D^{\mathrm{FH}}_{\max}$\\ 
  $D^{\mathrm{UB}}_{\max}$ & upper & [1,50] &  & \\
  $D^{\mathrm{UB}}_{15\%}$ & upper & [1,40] &  & $D^{\mathrm{UB}}_{15\%} < D^{\mathrm{UB}}_{\max}$\\
  $D^{\mathrm{UB}}_{40\%}$ & upper & [1,25] &  & $D^{\mathrm{UB}}_{40\%} < D^{\mathrm{UB}}_{15\%}$\\
  $D^{\mathrm{UB}}_{\mathrm{mean}}$ & upper & [1,20] &  &\\
\hline
\end{tabular}
\end{center}
\end{table}

% evaluation
% SAAS prior effectiveness evaluation
To investigate the effectiveness of the SAAS prior, we first compared the prediction
accuracy between standard GP and SAAS-GP on an identical data set. The data
set was created by optimizing one rectal cancer plan with planning parameters,
including 2000 entries of ($\theta, g(\theta)$) pair. The standard GP model and the
SAAS-GP model were trained on a subset of 100 samples, and another subset of unseen 100
samples was used to evaluate the prediction accuracies of both trained GP
models. This procedure was repeated 20 times for statistical evaluation.

With each patient, three automated treatment plans were obtained to compare with the original treatment plan (clinical): the
plan obtained by the random method and two BO plans obtained by GPEI and SAAS-BO, respectively.
The plan quality was evaluated based on target homogeneity index (HI)\cite{wambersie1999icru},
conformity index (CI)\cite{paddick2000simple}, and clinical goals in table \ref{tab:1}. The
aforementioned dosimetric indices of the four plans were compared and analyzed using Wilcoxon
signed-rank test, where the differences were considered statistically significant for $p < 0.05$.
In addition, the optimization efficiency among the three automated methods was
compared in terms of planning time and improvement per iteration.

To further compare the two BO methods, the length scales of the final fitted GP models
(standard GP and SAAS-GP) for each patient were analyzed to identify the sensitive optimization
parameters.
Following the automatic relevance determination (ARD) principle
\cite{williams2006gaussian}, the relative
importance of each parameter was quantified as the inverse of its corresponding length
scales. At last, the optimal planning parameters of three automated
planning methods were compared.

\section{Results}
\label{sec:3}
\subsection{Prediction accuracy}
\label{sec:3.1}

\begin{figure}[ht]
  \centering
  \captionsetup{justification=justified}
  \includegraphics[width=\textwidth]{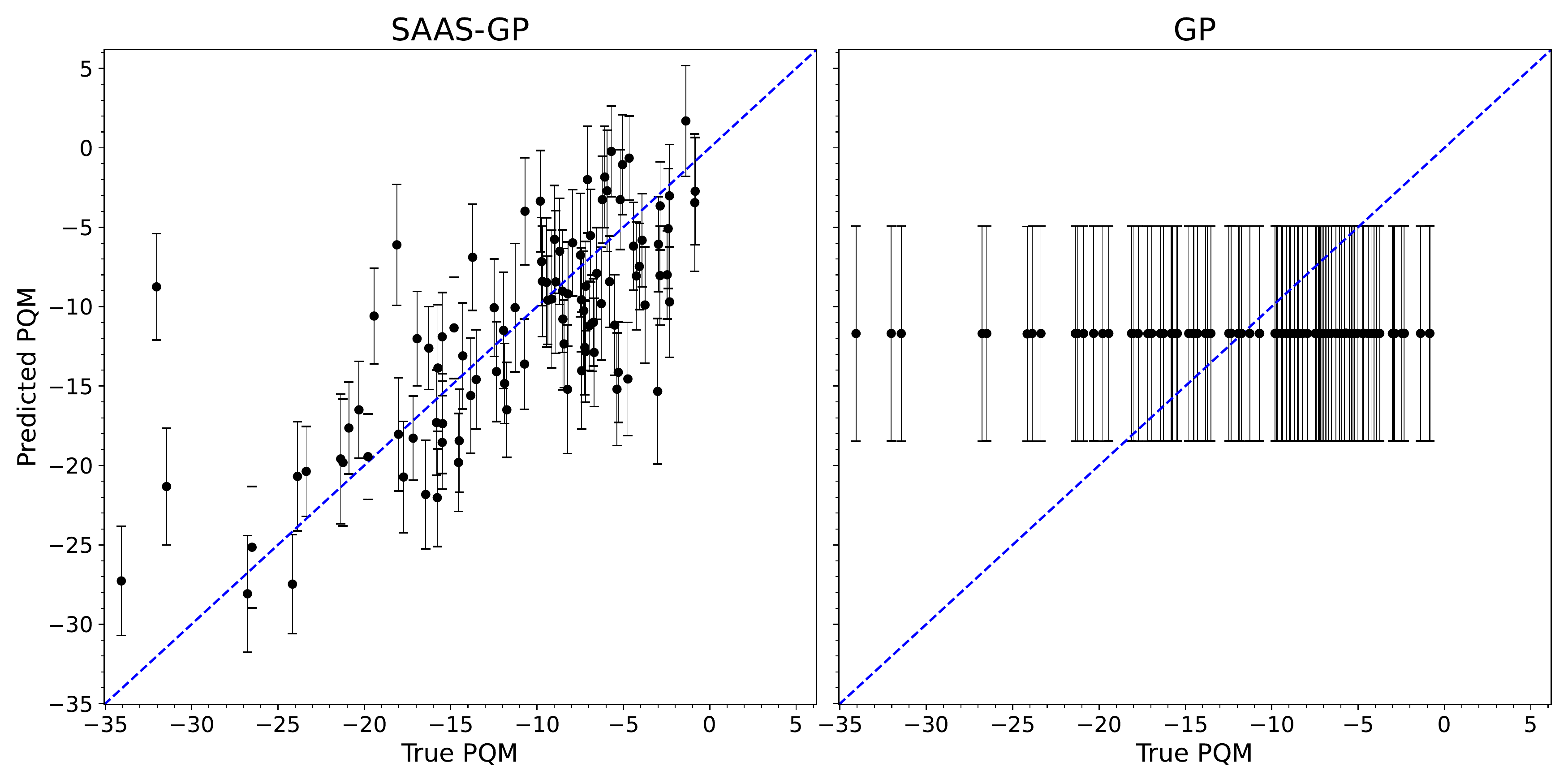}
  \caption{Prediction accuracies for both GP models on unseen sub-dataset,
    with the error bars representing 1 SD.}
  \label{fig:2}  
\end{figure}

Figure \ref{fig:2} demonstrates the predicted PQM score accuracy
comparison of the standard GP model and the SAAS-GP model, given identical
training and test data. Table \ref{tab:3} illustrates the
statistical diagnostics on the 20 repeated comparisons, including Pearson correlation
coefficient, Spearman's rank correlation coefficient, mean absolute percentage
error (MAPE)\cite{dodge2008concise}, and log likelihood on the test dataset. The
SAAS-GP model outperformed the GP model significantly in all the statistical measures.
The reasons for such differences in prediction accuracy were twofold:
1) The 100 random samples in the 34-D parameter space were sparsely
distributed; 2) There was a difference in the length scale assumptions for two GP
models - the standard GP model assumed equal importance of each dimension and
the learned length scales were relatively small, of the same magnitude. Since a
random testing point is highly likely to be far from the training samples (i.e.,
no similarity between the test input and the training samples can be exploited), the corresponding prediction
reverts to the mean over the training dataset. In contrast, the SAAS-GP model
assumed that only a limited number of dimensions were important (the length
scales of these dimensions are small), and the resulting predictions were more
meaningful for further optimization.

\begin{table}[ht]
\begin{center}
\arrayrulecolor{black}
\caption{Statistical diagnostics on the validation results for the standard GP and
  SAAS-GP model respectively.}
\label{tab:3}
\begin{tabular}{cccccc} 
\arrayrulecolor{black}\hline
 &  corr. coef. & rank corr. & MAPE & log likelihood \\
\arrayrulecolor{black}\hline
SAAS-GP & $0.79 \pm 0.04$ & $0.75 \pm 0.07$ & $0.66 \pm 0.28$& $-281.31 \pm 39.56$  \\
GP & $0.49 \pm 0.09 $ & $0.55 \pm 0.08$ & $1.43 \pm 0.57$ & $-330.89 \pm 45.96$ \\
\hline
\end{tabular}
\end{center}
\end{table}

\begin{figure}[ht]
  \centering
  \captionsetup{justification=justified}
  \includegraphics[width=\textwidth]{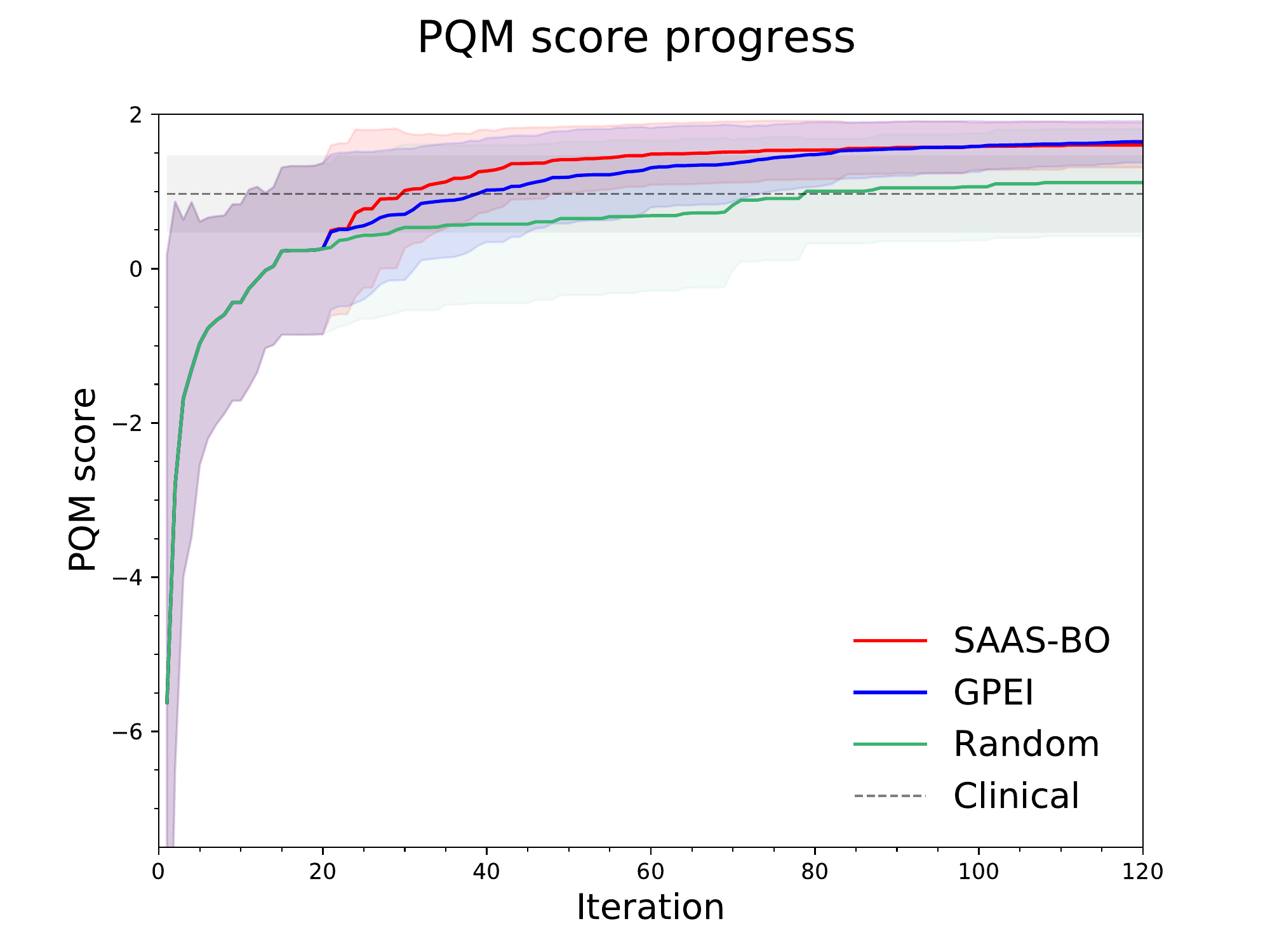}
  \caption{PQM score progress as a function of the iterations, along with the average
    PQM score of clinical treatment plans, where the color bands represent 1 SD.
  }
  \label{fig:3}
\end{figure}

Figure \ref{fig:3} demonstrates the best observed PQM scores per
iteration for random, GPEI, and SAAS-BO methods. Additionally, the
average PQM score of the clinical plans is also plotted for comparison.
At the end of the tuning phase, the average scores achieved by SAAS-BO and
GPEI surpassed that of the clinical plans with similar margins.
Meanwhile, the SAAS-BO spent fewer iterations to achieve convergence compared to
GPEI. Further, SAAS-BO achieved the smallest variance on the PQM
score over most of the iterations, indicating a more robust improvement than the GPEI
and random methods. % Adding optimization time here.
The optimization time of random, GPEI, and SAAS-BO are, respectively, 0.66 h ($\pm$
0.06 h), 3.16 h ($\pm$ 0.81 h), and 6.91 h ($\pm$ 0.43 h).

\subsection{Dosimetric comparison}
\label{sec:3.2}
\begin{figure}[ht]
  \centering
  \captionsetup{justification=justified}
  \includegraphics[width=\textwidth]{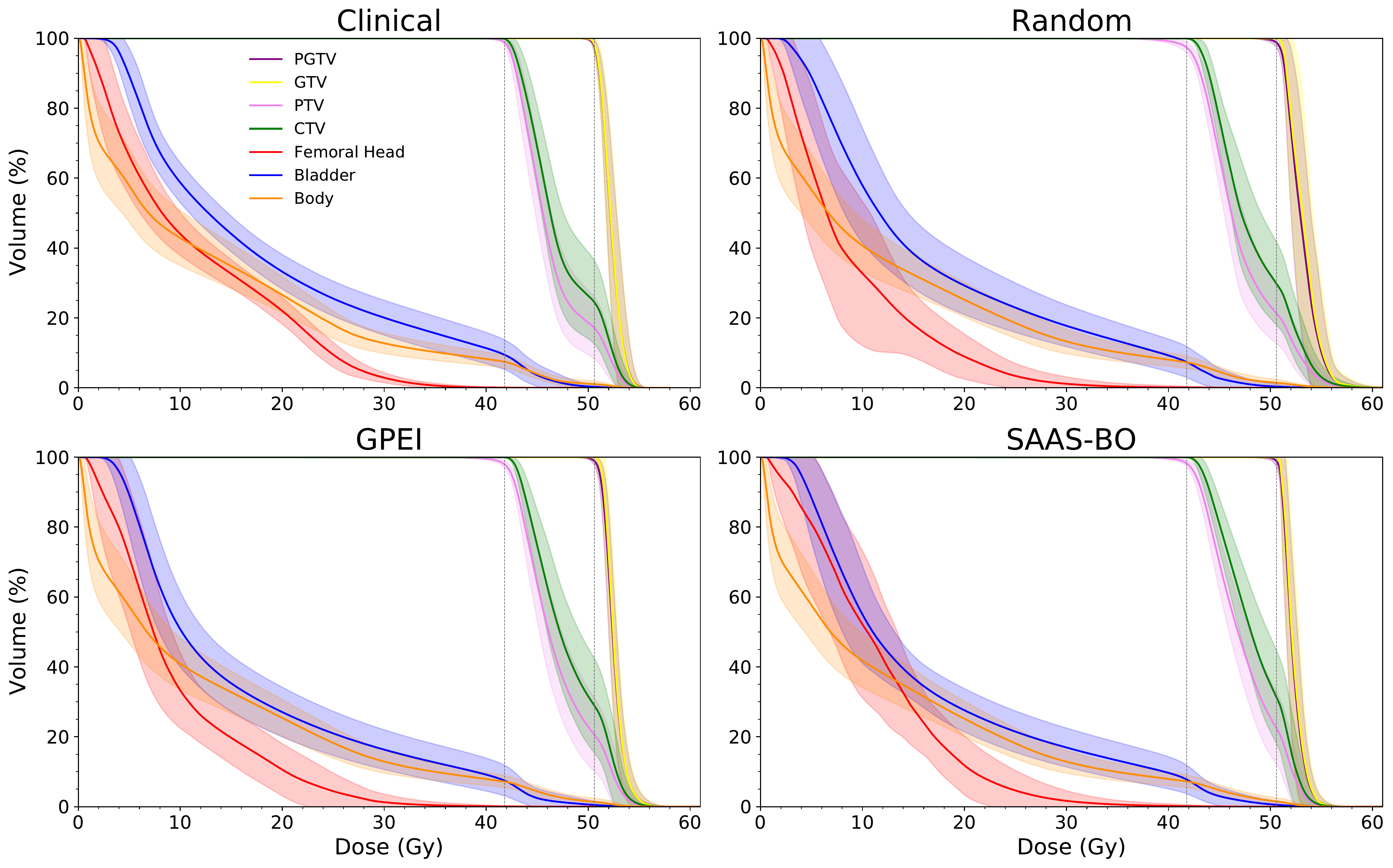}
  \caption{DVH comparison on the rectal cancer
    treatment plan cohort. The solid lines represent the means and the color
    bands represent 1 SD, and dose prescriptions were represented by gray dashed
    lines.}
  \label{fig:4}
\end{figure}

Figure \ref{fig:4} shows the mean DVH of the clinical, random, GPEI, and
SAAS-BO plans with corresponding standard deviations.
Compared with the clinical plans, all plans generated by three automated methods have shown some
improvement in OAR sparing, especially in the high dose region. For the GTV/PGTV
regions, the SAAS-BO plans have achieved comparable hot spot control with
the clinical plans, while the random and GPEI plans exhibited excessive high dose
tails.
Compared with the random and GPEI plans, the SAAS-BO plans have shown disadvantages in
the low-dose regions for the femoral head. 
Further dosimetric comparisons between clinical, random, GPEI, and SAAS-BO plans are
summarized in table \ref{tab:4}.
All plans have satisfied the prescribed requirements on
the dose coverages for the target ROIs.
The SAAS-BO plans demonstrated comparable hot spot control ($D_{2\%}$ in the PGTV) with the
clinical plans, significantly better than the GPEI and random plans ($p <
0.05$). Regarding the HI in the PGTV and the CI in the PTV, SAAS-BO plans achieved the best performances among the
four plans with significance ($p < 0.05$).
Related to the CI, both BO plans achieved better performances in the dose
spillage indicators ($p < 0.05$), indicating quicker dose
fall-off outside the PTV.

For the femoral head, significant reductions in $V_{20\mathrm{Gy}}$,
$V_{30\mathrm{Gy}}$ and $D_{\max}$ for three automatic plans have been observed
versus the clinical plans. The GPEI and random plans achieved significant reductions
in mean dose of the femoral head over the clinical plans.
For the bladder, major reductions in $V_{25\mathrm{Gy}}$ and mean dose have been
observed in SAAS-BO and GPEI plans, compared with the clinical plans. However, the maximum dose in SAAS-BO plans was higher than that of the clinical plans ($p < 0.05$).

\begin{table}[ht]
\renewcommand\arraystretch{1.25}
\begin{center}
\centering
\arrayrulecolor{black}
\caption{Dosimetric statistics among clinical plans, SAAS-BO plans, and GPEI
  plans. The Wilcoxon signed-rank test is used to compare plans,
  with significant values marked in bold (p \textless 0.05).}
\label{tab:4}
\scalebox{0.55}
{   
    \begin{tabular}{cccccccccccc} 
    \arrayrulecolor{black}\hline
    \multirow{2}{*}{Structure} & \multirow{2}{*}{Parameter} & \multirow{2}{*}{Clinical} & \multirow{2}{*}{SAAS-BO} & \multirow{2}{*}{GPEI} & \multirow{2}{*}{Random} & \multicolumn{6}{c}{p-value} \\ 
    \arrayrulecolor{black}\cline{7-12}
    % \hline
    &  &  &  &  &  & \makecell[c]{Clinical~vs.\\SAAS-BO} & \makecell[c]{Clinical~vs.\\GPEI} & \makecell[c]{Clinical~vs.\\Random} &  \makecell[c]{SAAS-BO~vs.\\GPEI} & \makecell[c]{SAAS-BO~vs.\\Random} & \makecell[c]{GPEI~vs.\\Random} \\ 
    \hline
      GTV & $D_{\min}$ & 50.40~(0.54)~ & 50.78~(0.38)~ & 50.92~(0.37)~ & 51.34~(0.98) & \textbf{0.009}~ & \textbf{0.001}~ &  \textbf{\textless0.001}~ & \textbf{0.036}~ &  0.013~ & 0.056~ \\
      \arrayrulecolor{black}\cline{2-6}
      CTV & $D_{\min}$ & 41.89~(0.46)~ & 42.20~(0.44)~ & 42.08~(0.35)~ & 42.10~(0.42) & 0.058~ & 0.083~ &  0.048~ & 0.518~ & 0.453~ & 1.000~ \\
      \arrayrulecolor{black}\cline{2-6}
    \multirow{4}{*}{PGTV} & $D_{98\%}$ & 50.72~(0.47)~ & 50.86~(0.14)~ & 50.97~(0.38)~ & 51.19~(1.04) & \textbf{0.048}~ & \textbf{0.015}~ & \textbf{0.036}~ & 0.475~  & 0.349~ & 0.546~ \\
     & $D_{95\%}$ & 50.99~(0.48)~ & 51.11~(0.35)~ & 51.26~(0.35)~ & 51.60~(0.94) & 0.133~ & \textbf{0.005}~  & \textbf{0.007}~ & 0.051~ & \textbf{0.019}~ & 0.189~ \\
                          & $D_{2\%}$ & 53.77~(0.71)~ & 53.98~(1.14)~ & 54.72~(0.85)~ & 55.66~(1.56) & 0.430~ & \textbf{\textless0.001}~ & \textbf{\textless0.001} & \textbf{0.044}~ & \textbf{\textless0.001} & \textbf{0.008} \\
      & HI & 0.06~(0.01)~ & 0.06~(0.02)~ & 0.07~(0.02)~ & 0.09~(0.03)& 0.956~ & \textbf{0.001}~ & \textbf{0.001}~ & \textbf{0.017}~  & \textbf{0.001} & 0.058~\\
    \arrayrulecolor{black}\cline{2-6}
    \multirow{5}{*}{PTV} & $D_{98\%}$ & 42.22~(0.47)~ & 41.96~(0.55)~ & 41.82~(0.41)~ & 41.46~(0.90) & 0.153~ & \textbf{0.011}~  & \textbf{0.004}~ & 0.123~& 0.058~ & 0.114~ \\
                               & $D_{95\%}$ & 42.68~(0.48)~ & 42.76~(0.46)~ & 42.57~(0.38)~ & 42.48~(0.58) & 0.522~ & 0.729~ & 0.202~ & 0.114~  & 0.076~ & 0.452~ \\
    & $R_{50\%}$ & 4.12~(0.25)~ & 3.91~(0.23)~ & 3.94~(0.17)~ & 3.94~(0.24) & \textbf{\textless0.001} & \textbf{0.001}~ & \textbf{0.001}~  & 0.261~ & 0.349~ & 0.985~\\
                               & $R_{90\%}$ & 1.50~(0.06)~ & 1.44~(0.04)~ & 1.44~(0.04)~ & 1.48~(0.10) & \textbf{\textless0.001}~ & \textbf{0.001}~ & 0.430~ & 0.546~  & 0.261~ & 0.143~\\
                               & CI & 0.73~(0.03)~ & 0.76~(0.02)~ & 0.76~(0.02)~ & 0.74~(0.04) & \textbf{\textless0.001}~ & \textbf{0.003}~ &  0.498~ &0.622~  & 0.114~ & 0.114~\\
      \arrayrulecolor{black}\cline{2-6}
    \multirow{4}{*}{Femoral Head} & $V_{20\mathrm{Gy}}$ & 22.00~(3.81) & 11.79~(8.44)~ & 10.57~(7.47)~ & 8.93~(6.44) & \textbf{\textless0.001} & \textbf{\textless0.001} & \textbf{\textless0.001} & 0.261~  & 0.177~ & 0.245~ \\
     & $V_{30\mathrm{Gy}}$ & 2.88~(1.44)~ & 2.07~(3.15)~ & 1.50~(2.45)~ & 1.63~(2.22) & \textbf{0.014} & \textbf{0.002} & \textbf{\textless0.001}~ & 0.784~  & \textbf{0.044}~ & 0.076~ \\
     & $D_{\max}$ & 41.59~(3.54)~ & 37.22~(8.84)~ & 36.38~(6.97)~ & 35.85~(8.14) & 0.058~ & \textbf{0.006}~ & \textbf{0.003}~ & 0.571~  & 0.430~ & 0.784~ \\
                               & $D_{\mathrm{mean}}$ & 11.37~(1.22) & 11.46~(2.27)~ & 9.51~(2.27)~ & 8.86~(2.73)  & 0.674~ & \textbf{\textless0.001}~ & \textbf{0.001}~ & \textbf{0.001}~  & \textbf{0.001}~ & 0.216~ \\
      \arrayrulecolor{black}\cline{2-6}
    \multirow{4}{*}{Bladder} & $V_{25\mathrm{Gy}}$ & 25.52~(4.99)~ & 21.46~(5.93)~ & 20.96~(6.34)~ & 22.92~(7.08) & \textbf{\textless0.001}~ & \textbf{\textless0.001} & \textbf{0.024}~ & 0.571~  & 0.090~ & \textbf{0.006}~ \\
     & $V_{45\mathrm{Gy}}$ & 3.63~(3.21)~ & 3.00~(2.65)~ & 2.57~(2.79)~ & 2.99~(2.93) & 0.202~ & \textbf{0.003}~ & \textbf{0.03}~& 0.123~ & 0.177~ & 0.701~\\
     & $D_{\max}$ & 50.56~(2.55)~ & 52.09~(3.62)~ & 50.46~(3.77)~ & 50.84~(4.69) & \textbf{0.053}~ & 0.596~ & 0.701~ & \textbf{0.004}~  & 0.154~ & 0.756~\\
     & $D_{\mathrm{mean}}$ & 17.58~(1.78)~ & 16.16~(2.18)~ & 15.71~(2.54)~ & 16.56~(3.10) & \textbf{0.002}~ & \textbf{\textless0.001}~ & 0.83~ & 0.330~  & 0.388~ & \textbf{0.014}~\\
    \hline
    \end{tabular}
}
\end{center}
\end{table}

\subsection{Parameter importance}
\begin{figure}[ht]
  \centering
  \captionsetup{justification=justified}
  \includegraphics[width=\textwidth]{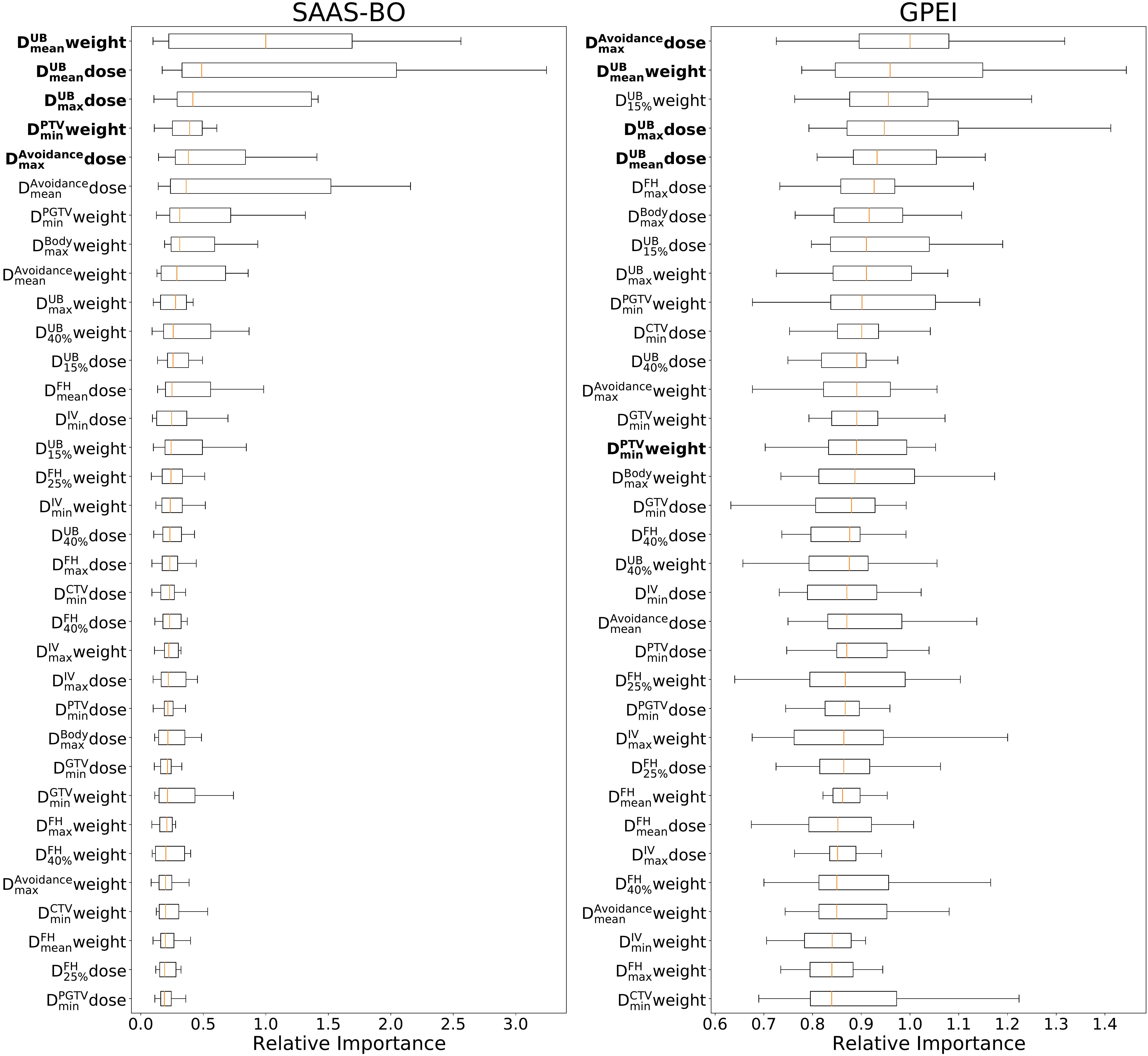}
  \caption{Relative importance box plot of the GP models related to
    the SAAS-BO and GPEI methods, averaged over the 20 rectal cancer treatment
    plans, normalized to the corresponding maximum values. The parameters were sorted according to their importance in descending order.
    The top 5 planning parameters in the SAAS-GP model
    were marked in bold, the orange line representing the median value.}
  \label{fig:5}
\end{figure}

Figure \ref{fig:5} summarizes the relative importance statistics of the 34 planning
parameters at the end of the BO (i.e., $T = 120$) for both GPEI
and SAAS-BO methods. 
We first note that the most important parameters between the two GP models were
quite consistent --- 4 out of the top 5 most important parameters in the SAAS-BO model
were in the top 5 in the GPEI model. Among the four common important parameters, three
were related to the bladder, and one to the avoidance structure. The detected important
parameters were consistent with the clinical experiences: both the contoured bladder and avoidance ROIs
were large, close to the targets, and often overlapped with the targets,
therefore affecting the PQM score significantly with small parameter changes.
Some differences can be though observed between the parameter importance of the two methods: The
magnitudes of the length scales from the GPEI methods were relatively uniform,
where the minimum relative importance was 83\% of the maximum relative
importance. Meanwhile, the minimum relative importance was 19\% of
the maximum relative importance in the SAAS-BO methods. This observation was
mainly due to the different prior between the SAAS-GP and GP models.
For both methods, although the important parameters were distinguishable from
the others, the distributions of their relative importance spanned large ranges,
indicating that these parameters were deemed not important in some cases.
The large variance might be caused by the correlated optimization parameters, e.g., the
mean dose objective and its weight for the bladder. On average, the target
objectives were less important than their corresponding weights,
because all parameter ranges were
normalized to [0, 1], while the varying ranges of the dose objectives for the
targets were much smaller than those of the objective weights.

\subsection{Optimal planning parameters}
\label{sec:3.3}
\begin{figure}[ht]
  \centering
  \captionsetup{justification=justified}
  \includegraphics[width=\textwidth]{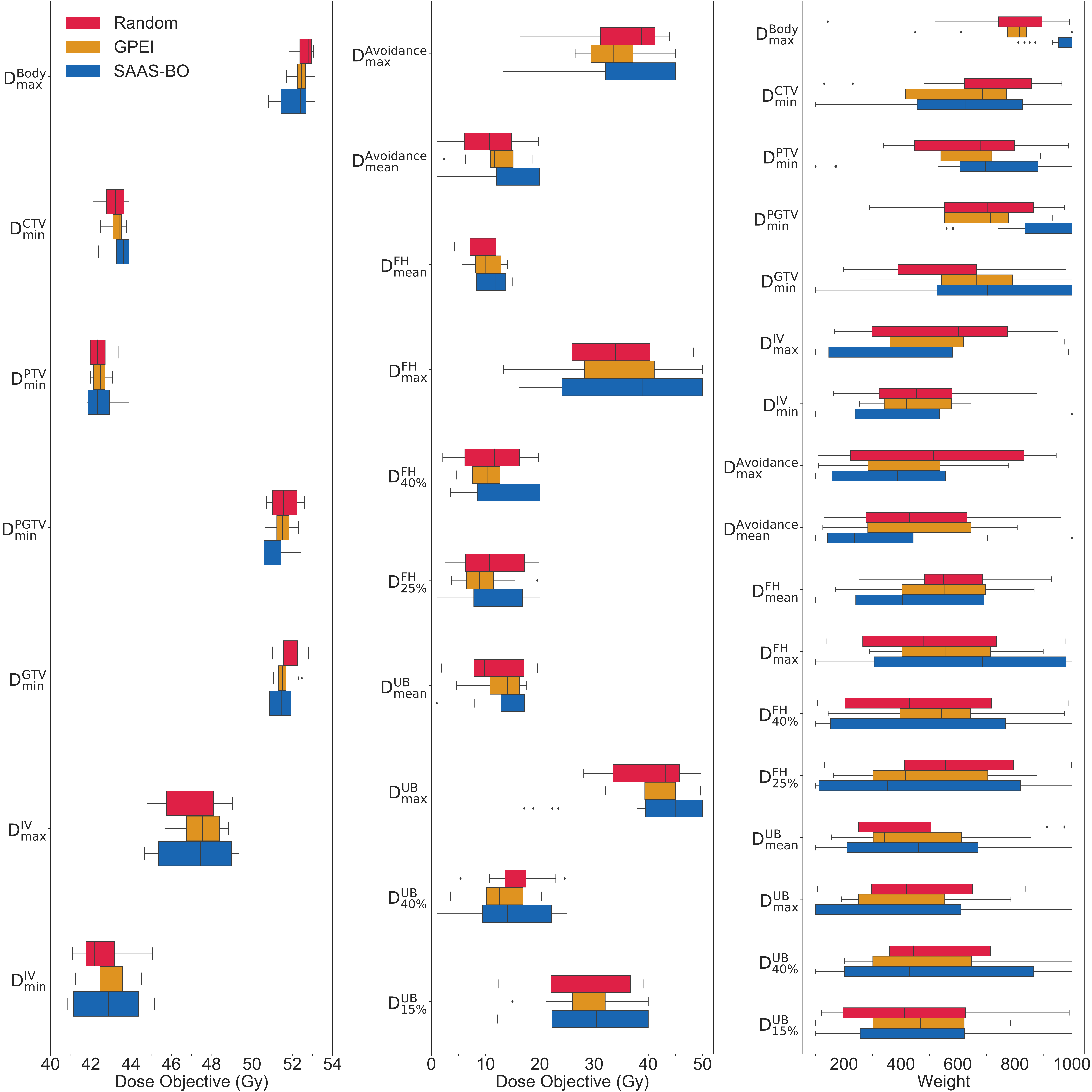}
  \caption{Optimal planning parameter distribution for 3 automated planning methods}
  \label{fig:6}
\end{figure}

We demonstrate the optimal parameter distributions for three automated planning
methods in figure \ref{fig:6}. In general, the optimal dose objectives for the OAR in SAAS-BO were set higher
on average compared to those in GPEI and random methods, indicating less focus
on the OAR sparing. However, The SAAS-BO methods emphasized on the
hot spot control in the target ROIs (i.e., $D_{\max}^{\mathrm{Body}}$ and its related
weight) compared to the other two methods. We note that most of the optimal
parameter distributions in SAAS-BO span larger ranges
compared to those in GPEI, indicating that GPEI performed more localized
searches in the parameter space.

\section{Discussion}
\label{sec:4}
Our method to automate the planning process belongs to the
automated iterative planning approach. The core idea is to incorporate
an additional global optimization loop so that optimal planning parameters can
be found during the outer iterations, which does not require prior plan data.
The advantages of this hyperparameter optimization approach are in
two main aspects:
\begin{itemize}
\item The clinical goals for the final plan quality evaluation do
  not need to be the same as the objectives used for inverse planning, and
  the clinical practice often falls into this scenario. The hyperparameter optimization 
  enables some flexibility in this regard, since a lot of
  dosimetric indices used in the plan quality evaluation might not be readily available
  as objectives in the optimization algorithm. This flexibility enables easier adaptation to
  clinical goal changes if patient-specific concern appears. 
\item The hyperparameter optimization was built on top of the TPS, acting as a self-contained module in the optimization workflow. The
  modular design reduces coupling with the inverse planning algorithm. Any
  update or feature change in the optimization can be adopted and inherited via
  hyperparameter optimization with less effort.
\end{itemize}

Among numerous global optimization methods, BO utilizes the GP regression to
model the underlying relation between the planning parameter space and PQM score.
Compared with the work of Maass et al.,\cite{maass2022hyperparameter} we further studied the BO
application in automated iterative planning with higher dimensionality. In addition, we
included both the dose limit and the corresponding weight into the adjustable
set, testing the automated planning approach in a more challenging scenario (with
minimum expert knowledge). 
We evaluated the performance of the recently proposed SAAS-BO
approach\cite{eriksson2021high} by comparing it with a GPEI
method and a random search method. Both BO methods achieved similar PQM
scores on the tested rectal cancer plans, significantly outperforming the clinical plans
in terms of the PQM score and the dosimetric indices (heterogeneity,
conformity, dose spillage, and DVH metrics).

Unlike other derivative-free optimization methods\cite{huang2022meta},
the BO approaches provide a GP regression model which learned the latent function
mapping the planning parameters to the PQM score. By analyzing the
length scales of the learned GP model, we observed that on average, both GP models
identified similar important planning parameters. In addition, the identified
parameters were consistent with clinical planning experiences.
However, large variance for the planning parameters can be observed, even for
the most important parameters, probably caused by the fact that the
intrinsic problem dimension is small, and the over-parameterization introduced
many correlated planning parameters. Although the over-parameterization
seemed excessive in our problem setting, this reflected to some extent the
common clinical practice. Our results demonstrated that the BO approaches were
able to outperform the clinical treatment plans, even under severe
over-parameterization conditions. 

We now compare the different behavior between the SAAS-BO and the GPEI
methods in detail. On random optimized datasets, the SAAS-GP exhibited significant
advantages over the standard GP in prediction accuracy, as shown in
figure \ref{fig:2}. This is consistent with the observation that SAAS-BO made a quicker PQM score improvement in the early tuning process than GPEI, as shown in
figure \ref{fig:3}. However, the SAAS-BO progress quickly reached bottlenecks
in the subsequent tuning process and underperformed the GPEI
in the end. In addition, we note that
SAAS-BO explored much larger search spaces than GPEI on the optimal planning parameter distributions. Combining these
two observations, we concluded that the reason for this behavior is that the
intrinsic important  parameters were not as sparse as the SAAS-BO prior assumed.
As a result, SAAS-BO tends to propose query input far to the current best input for most of the
``unimportant'' parameters (i.e., parameters with large length scales), and this search
pattern reduces the optimization efficiency at the end of the tuning process. On the contrary,
the explored region of the GPEI was very concentrated. This is because
the standard GP model overfitted and returned the trivial mean of the training dataset for most of the search space
(shown in figure \ref{fig:2}). The acquisition function proposed query input very
close to the current best input. In retrospect, a better optimization
strategy might be a hybrid one, i.e., performing the SAAS-BO to identify first the
important parameters and then switching to GPEI for more
localized searches.

The optimization time using BO approaches was relatively long, especially for the
SAAS-BO. In SAAS-BO, the GP model was learned in a fully Bayesian
manner, i.e., the Markov Chain Monte Carlo (MCMC) sampler was used to infer the
hyperparameter posterior density. The sampling approach increased the model fitting
time drastically compared to the gradient-based maximum likelihood approach. In
addition, we found that the numerous constraints in table \ref{tab:2}
significantly increased the evaluation time in the acquisition function. In an
ablation study, we removed all the constraints and performed hyperparameter optimization with the random, GPEI and SAAS-BO with identical settings (PQM score function, iteration budget,
etc.). The average optimization time of random, GPEI, and SAAS-BO are $0.41 \pm
0.01 $ h, $0.80 \pm 0.15$ h, and $2.86 \pm 0.39$ h, respectively. On average,
the optimization time was reduced to $\sim 40$\% of the time used in the
constrained setting. Conversely, the average final PQM scores without
constraints were reduced by 20\%, 6\%, and 5\% compared
with the average scores with constraints for the random, GPEI, and SAAS-BO
methods.
Among three automated planning methods, SAAS-BO benefited the least from the
constraints, indicating its intrinsic dimension reduction ability. The
corresponding PQM score progress plots and the DVHs were attached to the
supplementary material. To reduce further the hyperparameter optimization time,
techniques such as early stopping and parallel BO are to be considered.

Nevertheless, this work is subject to some limitations. The current BO
implementation considers the PQM score function as a black box, where the
GP model seemed overly flexible for some input dimensions. One way to
improve the optimization efficiency would be composite function BO
\cite{astudillo2019bayesian}, i.e., exposing the PQM score function structure to
the BO algorithm and modeling various terms in the score function by
multi-output GP.
Another direction would be to investigate the performance of the projection-based BO \cite{letham2020re} method
in automated treatment planning since many planning parameters appeared
strongly correlated, as shown in this work. In terms of clinical application,
we hope to extend the developed BO approach to other sites (e.g., H\&N, lungs,
etc.), investigating the dimensionality influences in more complex cases. 

\section{Conclusion}
\label{sec:5}
In this work, we compared the performances of two BO approaches (SAAS-BO and
GPEI) applied in automated treatment planning in high dimensional settings. The clinical
experiment was performed in the context of rectal cancer treatment planning.
We demonstrated that both BO methods were able to produce comparable or superior
plans compared with the clinical plans for all evaluated dosimetric indices.
Both methods were able to identify similar sensitive sets
of planning parameters with minimum expert knowledge. The developed
approach can be integrated with any TPS including
a scripting API, and is expected to ameliorate plan quality and reduce the
planning workload. 

\section*{Acknowledgements}
This work was supported in part by the Beijing Natural Science Foundation (No.
1202009, 1212011), National Key Research and Development Project (No.
2019YFF01014405), National Natural Science Foundation of China (No. 12005007),
Science Foundation of Peking University Cancer Hospital (No. 2021-14, KC2204, 2021-1).

\section*{References}
\addcontentsline{toc}{section}{\numberline{}References}
\vspace*{-20mm}

\bibliographystyle{./medphy.bst}
\bibliography{reference}

\begin{thebibliography}{10}

\bibitem{bijman2021pre}
R.~Bijman, A.~W. Sharfo, L.~Rossi, S.~Breedveld, and B.~Heijmen,
\newblock Pre-clinical validation of a novel system for fully-automated
  treatment planning,
\newblock Radiotherapy and Oncology {\bf 158}, 253--261 (2021).

\bibitem{craft2006approximating}
D.~L. Craft, T.~F. Halabi, H.~A. Shih, and T.~R. Bortfeld,
\newblock Approximating convex {P}areto surfaces in multiobjective radiotherapy
  planning,
\newblock Medical physics {\bf 33}, 3399--3407 (2006).

\bibitem{bortfeld2006imrt}
T.~Bortfeld,
\newblock {IMRT}: a review and preview,
\newblock Physics in Medicine \& Biology {\bf 51}, R363 (2006).

\bibitem{mcintosh2017fully}
C.~McIntosh, M.~Welch, A.~McNiven, D.~A. Jaffray, and T.~G. Purdie,
\newblock Fully automated treatment planning for head and neck radiotherapy
  using a voxel-based dose prediction and dose mimicking method,
\newblock Physics in Medicine \& Biology {\bf 62}, 5926 (2017).

\bibitem{fredriksson2012automated}
A.~Fredriksson,
\newblock Automated improvement of radiation therapy treatment plans by
  optimization under reference dose constraints,
\newblock Physics in Medicine \& Biology {\bf 57}, 7799 (2012).

\bibitem{babier2018inverse}
A.~Babier, J.~J. Boutilier, M.~B. Sharpe, A.~L. McNiven, and T.~C. Chan,
\newblock Inverse optimization of objective function weights for treatment
  planning using clinical dose-volume histograms,
\newblock Physics in Medicine \& Biology {\bf 63}, 105004 (2018).

\bibitem{chan2014generalized}
T.~C. Chan, T.~Craig, T.~Lee, and M.~B. Sharpe,
\newblock Generalized inverse multiobjective optimization with application to
  cancer therapy,
\newblock Operations Research {\bf 62}, 680--695 (2014).

\bibitem{boutilier2015models}
J.~J. Boutilier, T.~Lee, T.~Craig, M.~B. Sharpe, and T.~C. Chan,
\newblock Models for predicting objective function weights in prostate cancer
  {IMRT},
\newblock Medical Physics {\bf 42}, 1586--1595 (2015).

\bibitem{bokrantz2013distributed}
R.~Bokrantz,
\newblock Distributed approximation of {P}areto surfaces in multicriteria
  radiation therapy treatment planning,
\newblock Physics in Medicine \& Biology {\bf 58}, 3501 (2013).

\bibitem{xing1999optimization}
L.~Xing, J.~Li, S.~Donaldson, Q.~Le, and A.~Boyer,
\newblock Optimization of importance factors in inverse planning,
\newblock Physics in Medicine \& Biology {\bf 44}, 2525 (1999).

\bibitem{wu2003treatment}
C.~Wu, G.~H. Olivera, R.~Jeraj, H.~Keller, and T.~R. Mackie,
\newblock Treatment plan modification using voxel-based weighting factors/dose
  prescription,
\newblock Physics in Medicine \& Biology {\bf 48}, 2479 (2003).

\bibitem{huang2022meta}
C.~Huang, Y.~Nomura, Y.~Yang, and L.~Xing,
\newblock Meta-optimization for fully automated radiation therapy treatment
  planning,
\newblock Physics in Medicine \& Biology  (2022).

\bibitem{wang2021tree}
H.~Wang, R.~Wang, J.~Liu, J.~Zhang, K.~Yao, H.~Yue, Y.~Zhang, J.~You, and
  H.~Wu,
\newblock Tree-based exploration of the optimization objectives for automatic
  cervical cancer {IMRT} treatment planning,
\newblock The British Journal of Radiology {\bf 94}, 20210214 (2021).

\bibitem{zhang2011methodology}
X.~Zhang, X.~Li, E.~M. Quan, X.~Pan, and Y.~Li,
\newblock A methodology for automatic intensity-modulated radiation treatment
  planning for lung cancer,
\newblock Physics in Medicine \& Biology {\bf 56}, 3873 (2011).

\bibitem{shahriari2015taking}
B.~Shahriari, K.~Swersky, Z.~Wang, R.~P. Adams, and N.~De~Freitas,
\newblock Taking the human out of the loop: {A} review of {B}ayesian
  optimization,
\newblock Proceedings of the IEEE {\bf 104}, 148--175 (2015).

\bibitem{sambito2021strategies}
M.~Sambito and G.~Freni,
\newblock Strategies for improving optimal positioning of quality sensors in
  urban drainage systems for non-conservative contaminants,
\newblock Water {\bf 13}, 934 (2021).

\bibitem{turner2021bayesian}
R.~Turner, D.~Eriksson, M.~McCourt, J.~Kiili, E.~Laaksonen, Z.~Xu, and
  I.~Guyon,
\newblock Bayesian optimization is superior to random search for machine
  learning hyperparameter tuning: Analysis of the black-box optimization
  challenge 2020,
\newblock in {\em NeurIPS 2020 Competition and Demonstration Track}, pages
  3--26, PMLR, 2021.

\bibitem{yuan2019bayesian}
K.~Yuan, I.~Chatzinikolaidis, and Z.~Li,
\newblock Bayesian optimization for whole-body control of
  high-degree-of-freedom robots through reduction of dimensionality,
\newblock IEEE Robotics and Automation Letters {\bf 4}, 2268--2275 (2019).

\bibitem{lam2018advances}
R.~Lam, M.~Poloczek, P.~Frazier, and K.~E. Willcox,
\newblock Advances in Bayesian optimization with applications in aerospace
  engineering,
\newblock in {\em 2018 AIAA Non-Deterministic Approaches Conference}, page
  1656, 2018.

\bibitem{taasti2020automating}
V.~T. Taasti, L.~Hong, J.~S. Shim, J.~O. Deasy, and M.~Zarepisheh,
\newblock Automating proton treatment planning with beam angle selection using
  {B}ayesian optimization,
\newblock Medical Physics {\bf 47}, 3286--3296 (2020).

\bibitem{landers2018fully}
A.~C. Landers,
\newblock {\em Fully Automated Radiation Therapy Treatment Planning Through
  Knowledge-Based Dose Predictions},
\newblock University of California, Los Angeles, 2018.

\bibitem{wang2013bayesian}
Z.~Wang et~al.,
\newblock Bayesian Optimization in High Dimensions via Random Embeddings.,
\newblock in {\em IJCAI}, pages 1778--1784, Citeseer, 2013.

\bibitem{maass2022hyperparameter}
K.~Maass, A.~Aravkin, and M.~Kim,
\newblock A hyperparameter-tuning approach to automated inverse planning,
\newblock Medical Physics  (2022).

\bibitem{Dong2013}
P.~Dong, P.~Lee, D.~Ruan, T.~Long, E.~Romeijn, Y.~Yang, D.~Low, M.~Kupelian,
  Patrick, and K.~Sheng,
\newblock 4$\pi$ Non-Coplanar Liver {SBRT}: A Novel Delivery Technique,
\newblock International journal of Radiation Oncology* Biology* Physics {\bf
  85}, 1360--1366 (2013).

\bibitem{lu2007reduced}
R.~Lu, R.~J. Radke, L.~Happersett, J.~Yang, C.-S. Chui, E.~Yorke, and
  A.~Jackson,
\newblock Reduced-order parameter optimization for simplifying prostate {IMRT}
  planning,
\newblock Physics in Medicine \& Biology {\bf 52}, 849 (2007).

\bibitem{li2012preoperative}
J.-l. Li et~al.,
\newblock Preoperative concomitant boost intensity-modulated radiotherapy with
  oral capecitabine in locally advanced mid-low rectal cancer: a phase {II}
  trial,
\newblock Radiotherapy and Oncology {\bf 102}, 4--9 (2012).

\bibitem{williams2006gaussian}
C.~K. Williams and C.~E. Rasmussen,
\newblock {\em Gaussian processes for machine learning},
\newblock MIT press Cambridge, MA, 2006.

\bibitem{jones1998efficient}
D.~R. Jones, M.~Schonlau, and W.~J. Welch,
\newblock Efficient global optimization of expensive black-box functions,
\newblock Journal of Global Optimization {\bf 13}, 455--492 (1998).

\bibitem{bratley1988algorithm}
P.~Bratley and B.~L. Fox,
\newblock Algorithm 659: Implementing Sobol's quasirandom sequence generator,
\newblock ACM Transactions on Mathematical Software (TOMS) {\bf 14}, 88--100
  (1988).

\bibitem{eriksson2021high}
D.~Eriksson and M.~Jankowiak,
\newblock {H}igh-dimensional {B}ayesian optimization with sparse axis-aligned
  subspaces,
\newblock in {\em Uncertainty in Artificial Intelligence}, pages 493--503,
  PMLR, 2021.

\bibitem{wang2016bayesian}
Z.~Wang, F.~Hutter, M.~Zoghi, D.~Matheson, and N.~de~Feitas,
\newblock {B}ayesian optimization in a billion dimensions via random
  embeddings,
\newblock Journal of Artificial Intelligence Research {\bf 55}, 361--387
  (2016).

\bibitem{balandat2020botorch}
M.~Balandat, B.~Karrer, D.~Jiang, S.~Daulton, B.~Letham, A.~G. Wilson, and
  E.~Bakshy,
\newblock {B}o{T}orch: a framework for efficient {M}onte-{C}arlo {B}ayesian
  optimization,
\newblock Advances in Neural Information Processing Systems {\bf 33},
  21524--21538 (2020).

\bibitem{vassiliev2010validation}
O.~N. Vassiliev, T.~A. Wareing, J.~McGhee, G.~Failla, M.~R. Salehpour, and
  F.~Mourtada,
\newblock Validation of a new grid-based Boltzmann equation solver for dose
  calculation in radiotherapy with photon beams,
\newblock Physics in Medicine \& Biology {\bf 55}, 581 (2010).

\bibitem{letham2020re}
B.~Letham, R.~Calandra, A.~Rai, and E.~Bakshy,
\newblock Re-examining linear embeddings for high-dimensional {B}ayesian
  optimization,
\newblock Advances in Neural Information Processing Systems {\bf 33},
  1546--1558 (2020).

\bibitem{astudillo2019bayesian}
R.~Astudillo and P.~Frazier,
\newblock {B}ayesian optimization of composite functions,
\newblock in {\em International Conference on Machine Learning}, pages
  354--363, PMLR, 2019.

\bibitem{2019Safety}
A.~Bezjak, R.~Paulus, L.~E. Gaspar, R.~D. Timmerman, W.~L. Straube, W.~F. Ryan,
  Y.~I. Garces, A.~T. Pu, A.~K. Singh, and G.~M. Videtic,
\newblock Safety and Efficacy of a Five-Fraction Stereotactic Body Radiotherapy
  Schedule for Centrally Located Non–Small-Cell Lung Cancer: {NRG}
  Oncology/{RTOG} 0813 Trial,
\newblock Journal of Clinical Oncology  (2019).

\bibitem{hong2015nrg}
T.~S. Hong et~al.,
\newblock {NRG} {Oncology Radiation Therapy Oncology Group} 0822: a phase 2
  study of preoperative chemoradiation therapy using intensity modulated
  radiation therapy in combination with capecitabine and oxaliplatin for
  patients with locally advanced rectal cancer,
\newblock International Journal of Radiation Oncology* Biology* Physics {\bf
  93}, 29--36 (2015).

\bibitem{dodge2008concise}
Y.~Dodge,
\newblock {\em The concise encyclopedia of statistics},
\newblock Springer Science \& Business Media, 2008.

\bibitem{shen2020operating}
C.~Shen, D.~Nguyen, L.~Chen, Y.~Gonzalez, R.~McBeth, N.~Qin, S.~B. Jiang, and
  X.~Jia,
\newblock Operating a treatment planning system using a deep-reinforcement
  learning-based virtual treatment planner for prostate cancer
  intensity-modulated radiation therapy treatment planning,
\newblock Medical physics {\bf 47}, 2329--2336 (2020).

\bibitem{sobotta2008tools}
B.~Sobotta, M.~S{\"o}hn, M.~P{\"u}tz, and M.~Alber,
\newblock Tools for the analysis of dose optimization: III. Pointwise
  sensitivity and perturbation analysis,
\newblock Physics in Medicine \& Biology {\bf 53}, 6337 (2008).

\bibitem{alber2002tools}
M.~Alber, M.~Birkner, and F.~N{\"u}sslin,
\newblock Tools for the analysis of dose optimization: II. Sensitivity
  analysis,
\newblock Physics in Medicine \& Biology {\bf 47}, N265 (2002).

\bibitem{bokrantz2013multicriteria}
R.~Bokrantz,
\newblock {\em Multicriteria optimization for managing tradeoffs in radiation
  therapy treatment planning},
\newblock PhD thesis, KTH Royal Institute of Technology, 2013.

\bibitem{paddick2000simple}
I.~Paddick,
\newblock A simple scoring ratio to index the conformity of radiosurgical
  treatment plans,
\newblock Journal of neurosurgery {\bf 93}, 219--222 (2000).

\bibitem{wambersie1999icru}
A.~Wambersie,
\newblock {ICRU} {R}eport 62, prescribing, recording and reporting photon beam
  therapy (supplement to {ICRU} {R}eport 50),
\newblock {ICRU} News  (1999).

\bibitem{bakshy2018ae}
E.~Bakshy, L.~Dworkin, B.~Karrer, K.~Kashin, B.~Letham, A.~Murthy, and
  S.~Singh,
\newblock {AE}: {A} domain-agnostic platform for adaptive experimentation,
\newblock in {\em Workshop on Systems for ML}, pages 1--8, 2018.

\bibitem{huang2021fully}
C.~Huang, Y.~Yang, and L.~Xing,
\newblock Fully automated noncoplanar radiation therapy treatment planning,
\newblock Medical Physics {\bf 48}, 7439--7449 (2021).

\bibitem{hong2008multicriteria}
T.~S. Hong, D.~L. Craft, F.~Carlsson, and T.~R. Bortfeld,
\newblock Multicriteria optimization in intensity-modulated radiation therapy
  treatment planning for locally advanced cancer of the pancreatic head,
\newblock International Journal of Radiation Oncology* Biology* Physics {\bf
  72}, 1208--1214 (2008).

\bibitem{breedveld2007novel}
S.~Breedveld, P.~R. Storchi, M.~Keijzer, A.~W. Heemink, and B.~J. Heijmen,
\newblock A novel approach to multi-criteria inverse planning for {IMRT},
\newblock Physics in Medicine \& Biology {\bf 52}, 6339 (2007).

\bibitem{fan2019automatic}
J.~Fan, J.~Wang, Z.~Chen, C.~Hu, Z.~Zhang, and W.~Hu,
\newblock {A}utomatic treatment planning based on three-dimensional dose
  distribution predicted from deep learning technique,
\newblock Medical Physics {\bf 46}, 370--381 (2019).

\end{thebibliography}
\end{document}

\title{Supplementary Material}
\maketitle
\setlength{\parindent}{2ex}
This document contains the detailed model specifications and supplementary results
to accompany the manuscript named: ``High-dimensional Automated Radiation
Treatment Planning via Bayesian Optimization''.
\section{GP model specification}
Here the full specifications for the SAAS-GP and GP
regression models are listed, respectively.

GP (used in the GPEI):
\begin{eqnarray*}
  \mathrm{kernel\ scale}\ &\sigma_k^2 &\sim \mathrm{Gamma}(2, 0.15),\nonumber\\
  \mathrm{length\ scale} &l_i & \sim \mathrm{Gamma}(3, 6)\ \mathrm{for}\ i = 1,
                                \ldots, s,\nonumber \\
  \mathrm{distance}: &r(\theta, \theta')^2&= \sum \limits_{i=1}^s \left ((w_i
                                            - w_i') / l_i\right)^2,\nonumber\\
  \mathrm{covariance}: &k(\theta, \theta')&= \sigma_k^2(1 + \sqrt{5}r + \frac{5}{3}r^2 )\exp(-\sqrt{5}r),\nonumber\\
  \mathrm{function\ value}: & \hat{g}(\theta) &\sim \mathcal{GP}(0,
                                                k(\theta, \theta')),
                                                \nonumber \\
  \mathrm{observation}: &  g(\theta) &\sim
                                       \mathcal{N}(\hat{g}(\theta),
                                       \sigma^2),
\end{eqnarray*}

SAAS-GP (used in the SAAS-BO): 
\begin{eqnarray*}
  \mathrm{kernel\ scale} &\sigma_k^2 &\sim \mathrm{Gamma}(2,
                                          0.15),\nonumber\\
  \mathrm{shrinkage\ scale} &\tau &\sim \mathcal{HC}(0.1), \nonumber \\
  \mathrm{inverse\ squared\ length\ scale}\ &\rho_i &\sim \mathcal{HC}(\tau)\ \mathrm{for}\ i
                                          = 1, \ldots, s,\nonumber \\
  \mathrm{length\ scale} &l_i &  = \sqrt{1/\rho_i}, \nonumber\\
  \mathrm{distance}: &r(\theta, \theta')^2&= \sum \limits_{i=1}^s \left ((w_i
                                            - w_i') / l_i\right)^2,\nonumber\\
  \mathrm{covariance}: &k(\theta, \theta')&= \sigma_k^2(1 + \sqrt{5}r + \frac{5}{3}r^2 )\exp(-\sqrt{5}r),\nonumber\\
  \mathrm{function\ value}: & \hat{g}(\theta) &\sim \mathcal{GP}(0,
                                                k(\theta, \theta')),
                                                \nonumber \\
  \mathrm{observation}: &  g(\theta) &\sim
                                       \mathcal{N}(\hat{g}(\theta),
                                       \sigma^2) \\
  \mathrm{constraints}: & & l_i\textgreater0,  \rho_i\textgreater0, \tau\textgreater0
\end{eqnarray*}
where $\mathrm{Gamma}$ denotes the gamma distribution and $\mathcal{HC}$ denotes
the half-Cauchy distribution. One can notice the major difference between the
two model lies in the length scale priors. Such difference in
the priors leads to different inference strategies to infer the posterior density
of the GP hyperparameters $\psi=\{\sigma_k,
l_1, \ldots, l_s\}$ given the data.

According to the Bayes' theorem, the posterior distribution of the $\psi$ for the
standard GP model can
be expressed as
\begin{equation}
  p(\psi \vert \Theta, \mathbf{g}) \propto p(\mathbf{g} \vert \Theta, \psi) p(\psi),
  \label{eq:1}
\end{equation}
where $\Theta = \{\theta_i\}_{i=1}^N$ denotes the $N$ set of planning parameters
, $\mathbf{g} = \{g(\theta_i)\}_{i=1}^N$ the corresponding observed
PQM scores, and $p(\mathbf{g} \vert \Theta, \psi)$ the marginal likelihood,
which can be calculated in closed form
\begin{equation}
  p(\mathbf{g} \vert \Theta, \psi) \sim \mathcal{N}(\mathbf{g}, K_\psi + \sigma^2 I),
\end{equation}
where $K_{\psi}$ denotes the $N \times N$ covariance matrix governed by the
hyperparameters  $\psi$. The rhs product in eq.\ref{eq:1} can be computed in closed
form, and maximizing $p(\psi \vert \Theta, \mathbf{g})$ with regard to
$\psi$ yields the MAP estimate of $\psi$.
In the SAAS-GP, the length scale prior is hierarchical,
where the shrinkage scale $\tau$ is governed by another half-Cauchy prior
distribution. The joint posterior of $\tau$ and $\psi$ is expressed as
\begin{equation}
  p(\tau, \psi \vert \Theta, \mathbf{g}) \propto p(\mathbf{g}) p(\psi \vert \tau) p(\tau),
  \label{eq:3}
\end{equation}
where $p(\tau)$ denotes the prior distribution for $\tau$, and $p(\psi \vert
\tau)$ the prior for $\psi$ governed by $\tau$. A straightforward way to infer
the joint posterior distribution with the hierarchical model in eq.\ref{eq:3} is
to perform MCMC sampling. In this work, the No-U-Turn Sampler was used to
perform inference, and the resulting posterior distribution of the
hyperparameters was represented by multiple samples $\{{\psi_i}\}_{i=1}^n$. As a result, the MCMC
-based inference is much slower than the MAP-based inference. 

\section{Ablation study: BO without constraints}
\begin{figure}[H]
  \centering
  \captionsetup{justification=justified}
  \includegraphics[width=\textwidth]{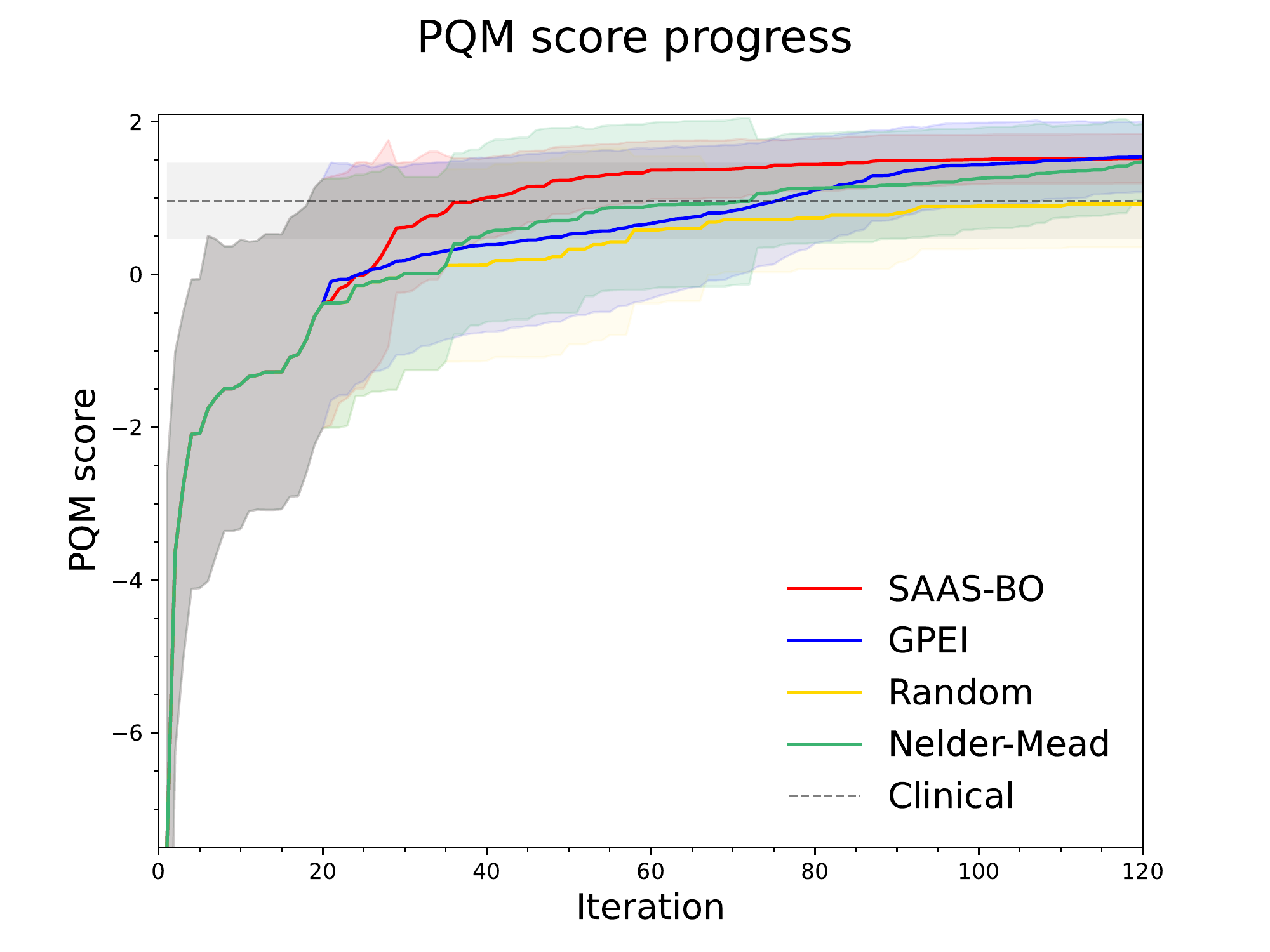}
  \caption{automated planning PQM scores without parameter constraints as a function of the iterations, along with the average PQM score of clinical treatment plans, where the color bands represent 1 SD.}
  \label{fig:1}
\end{figure}

\begin{figure}[H]
  \centering
  \captionsetup{justification=justified}
  \includegraphics[width=\textwidth]{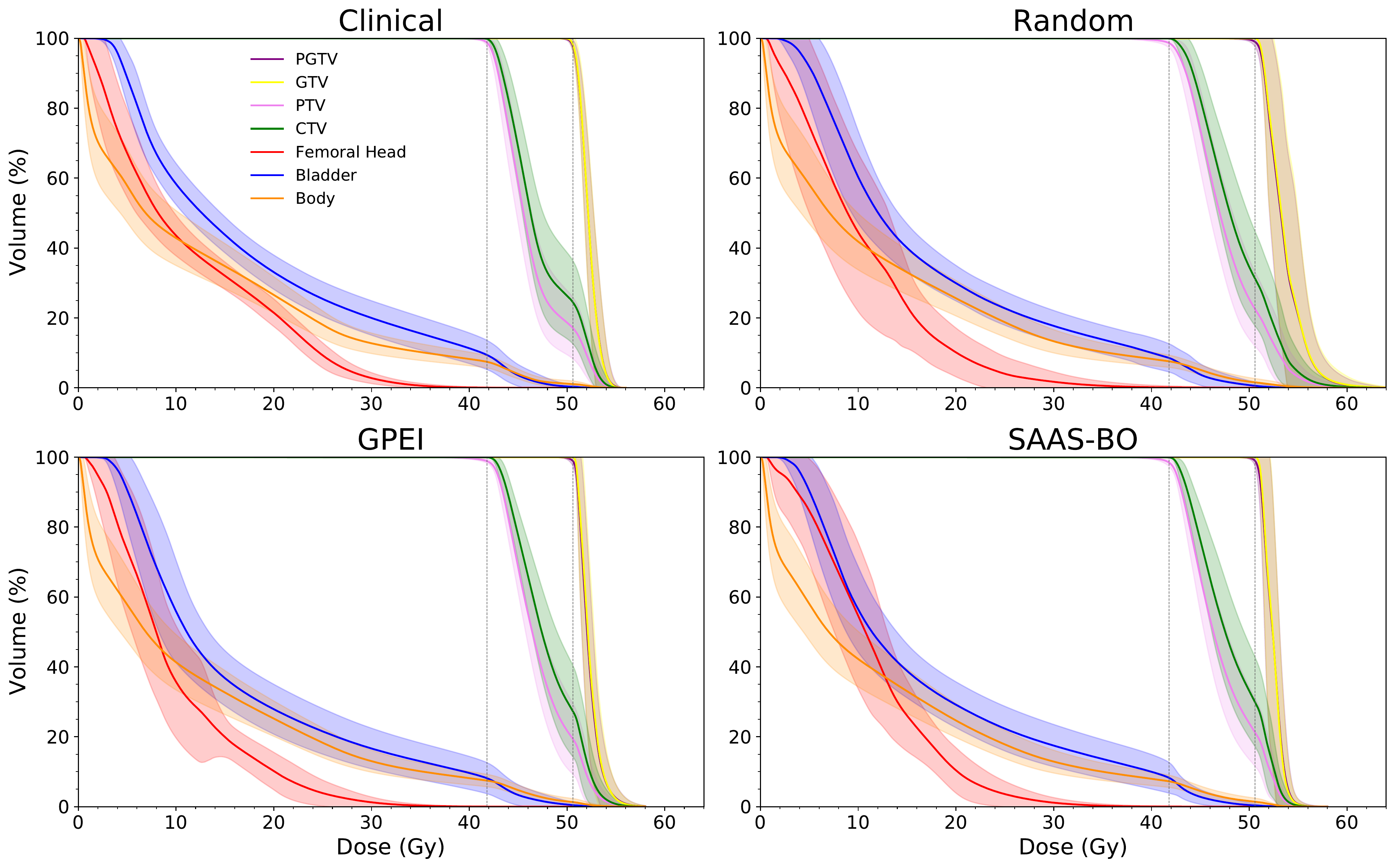}
  \caption{DVH comparison on the rectal cancer treatment plan cohort without
    constraints in the search space. The solid lines represent the means and the color bands represent 1 SD, and dose prescriptions were represented by gray dashed lines.}
  \label{fig:2}
\end{figure}

As mentioned in the discussion, we performed an ablation study based on
identical settings (PQM score function, iteration budget, etc.) to the previous
experiment, except for removing all constraints on the adjustable parameters.
The results obtained by four auto-planning methods (random, GPEI, and SAAS-BO,
Nelder-Mead) are shown in figure \ref{fig:1} and \ref{fig:2}. Compared with the
progress with constraints in the manuscript, one can observe that the early
progress advantage for the SAAS-BO is more obvious in the no-constraint
experiment. 

\section{DVH data for the random method}

\begin{figure}[H]
  \centering
  \captionsetup{justification=justified}
  \includegraphics[width=\textwidth]{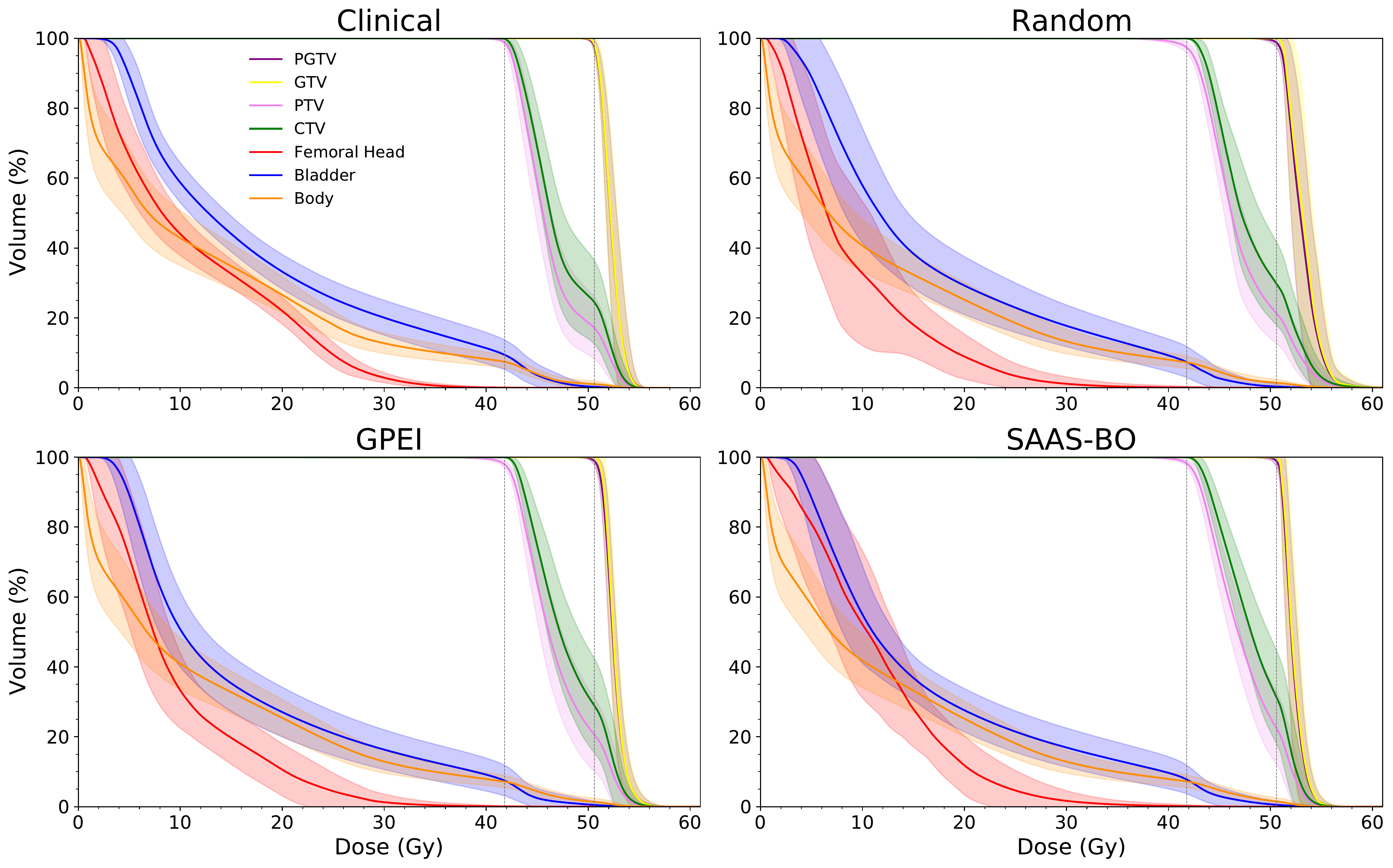}
  \caption{DVH comparison on the rectal cancer treatment plan cohort with
    constraints in the search space. The solid lines represent the means and the color bands represent 1 SD, and dose prescriptions were represented by gray dashed lines.}
  \label{fig:3}
\end{figure}

\setlength{\parindent}{2ex}
\indent
As mentioned in the methods, DVH for the random method are shown in
figure \ref{fig:3}. The results of clinical, GPEI, and SAAS-BO methods are identical to the DVH results in the main text\\
\indent For the GTV and PGTV, the random plans exhibited excessive high dose
tails and poor hot spot control. However, the random plans have shown good control in the low-dose region for the femoral head.

\section{The impact of different clinical goal priorities}
 \indent We designed score functions with three different sets of clinical goal
 priorities, as is shown in table \ref{tab:1}, where the $\mathrm{PQM_{target}}$
 is identical to the priority settings reported in the manuscript. In this
 set of priority settings, the target was given the most important priority and
 the OAR, dose spillage were considered with less importance.  
 In $\mathrm{PQM_{equal}}$, all scoring terms are set equally important,
 while $\mathrm{PQM_{OAR}}$ sets the terms of OAR and dose spillage to the top
 tier and the target was set less important.

\begin{figure}[H]
  \centering
  \captionsetup{justification=justified}
  \includegraphics[width=\textwidth]{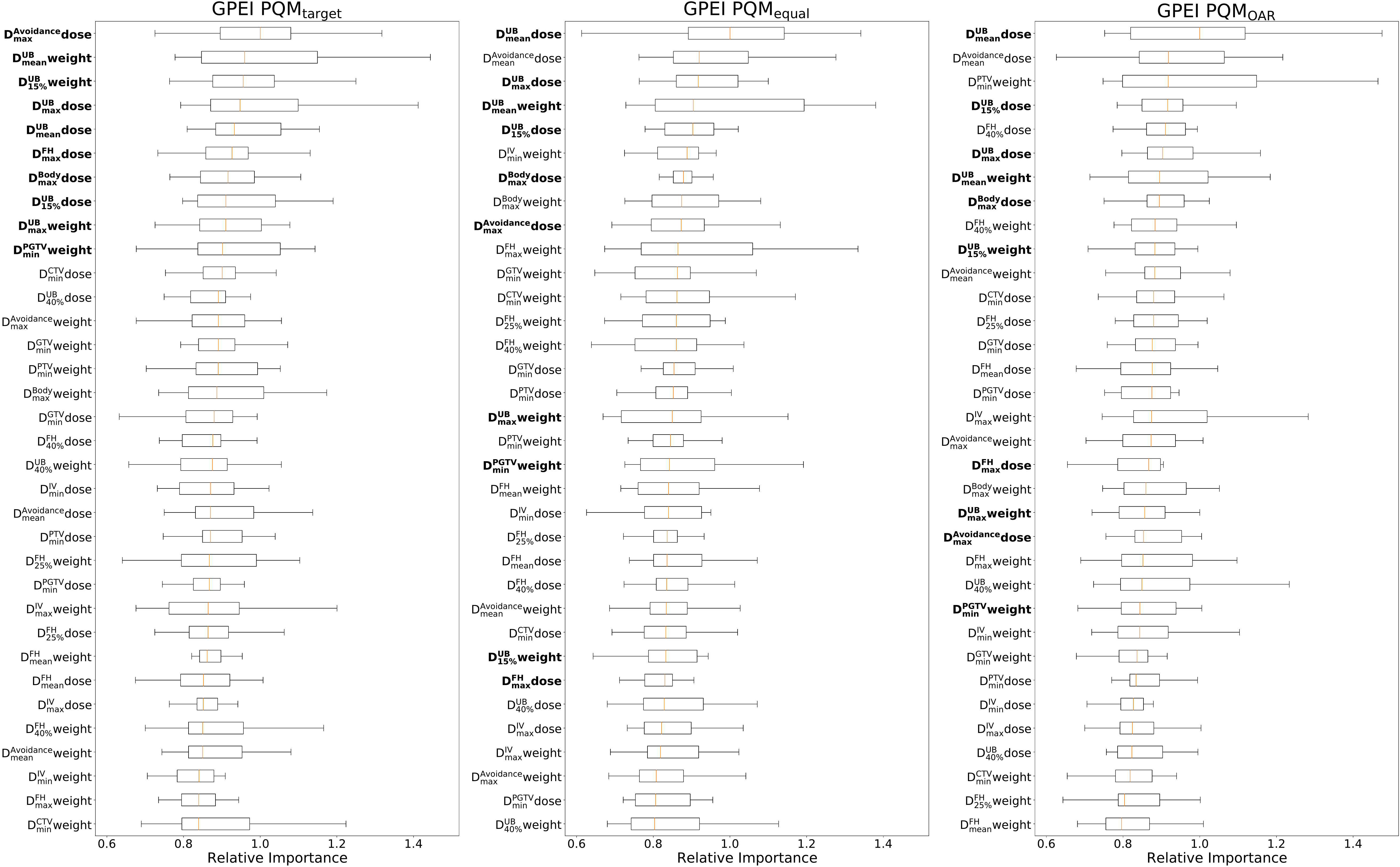}
  \caption{Relative importance box plot of GPEI methods based on different
    clinical priorities,
    normalized to the corresponding maximum of 34 median importance values over 20 cases. The parameters were sorted
according to their importance in descending order. The top 10 planning parameters in the GPEI $\mathrm{PQM_{target}}$
model were marked in bold, the orange line representing the median value}
  \label{fig:4}
\end{figure}

\begin{figure}[H]
  \centering
  \captionsetup{justification=justified}
  \includegraphics[width=\textwidth]{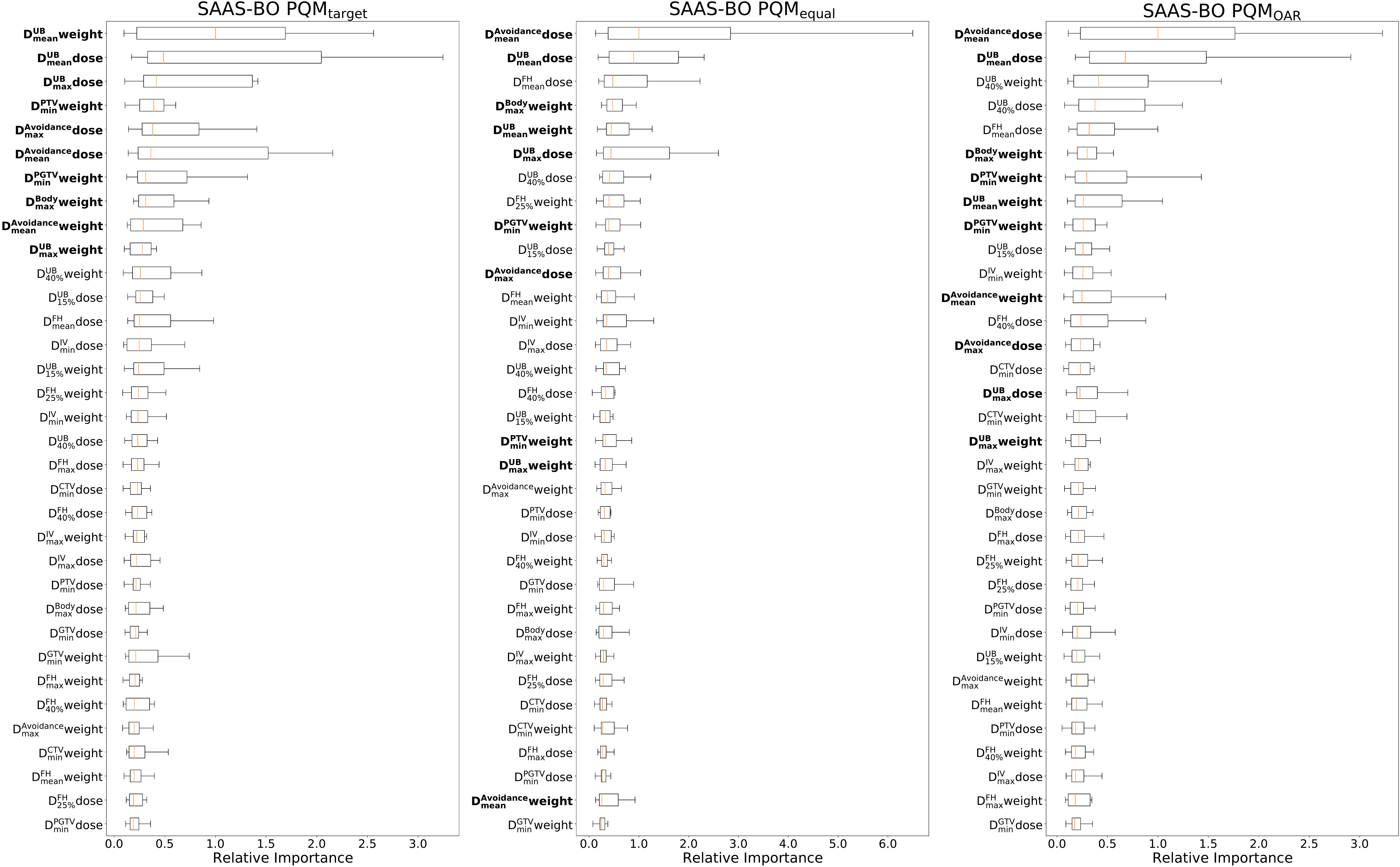}
  \caption{Relative importance box plot of SAAS-BO methods based on different
    clinical priority sets.}
  \label{fig:5}
\end{figure}

 \indent The relative importance statistics of the 34 planning parameters for
 GPEI and SAAS-BO methods based on different score functions were shown in
 figure \ref{fig:4} and \ref{fig:5}, respectively. We made the following
 qualitative observations. When the terms of targets are at the top priority, the
 relative importance of the parameters corresponding to the target area (PTV and
 PGTV) and its surrounding structures (avoidance and bladder) are ranked higher.
 As the priority of OAR and dose spillage increased, the relative importance of parameters that are less relevant to the target increased in the ranking, such as the terms of femoral head.

\begin{figure}[H]
  \centering
  \captionsetup{justification=justified}
  \includegraphics[width=\textwidth]{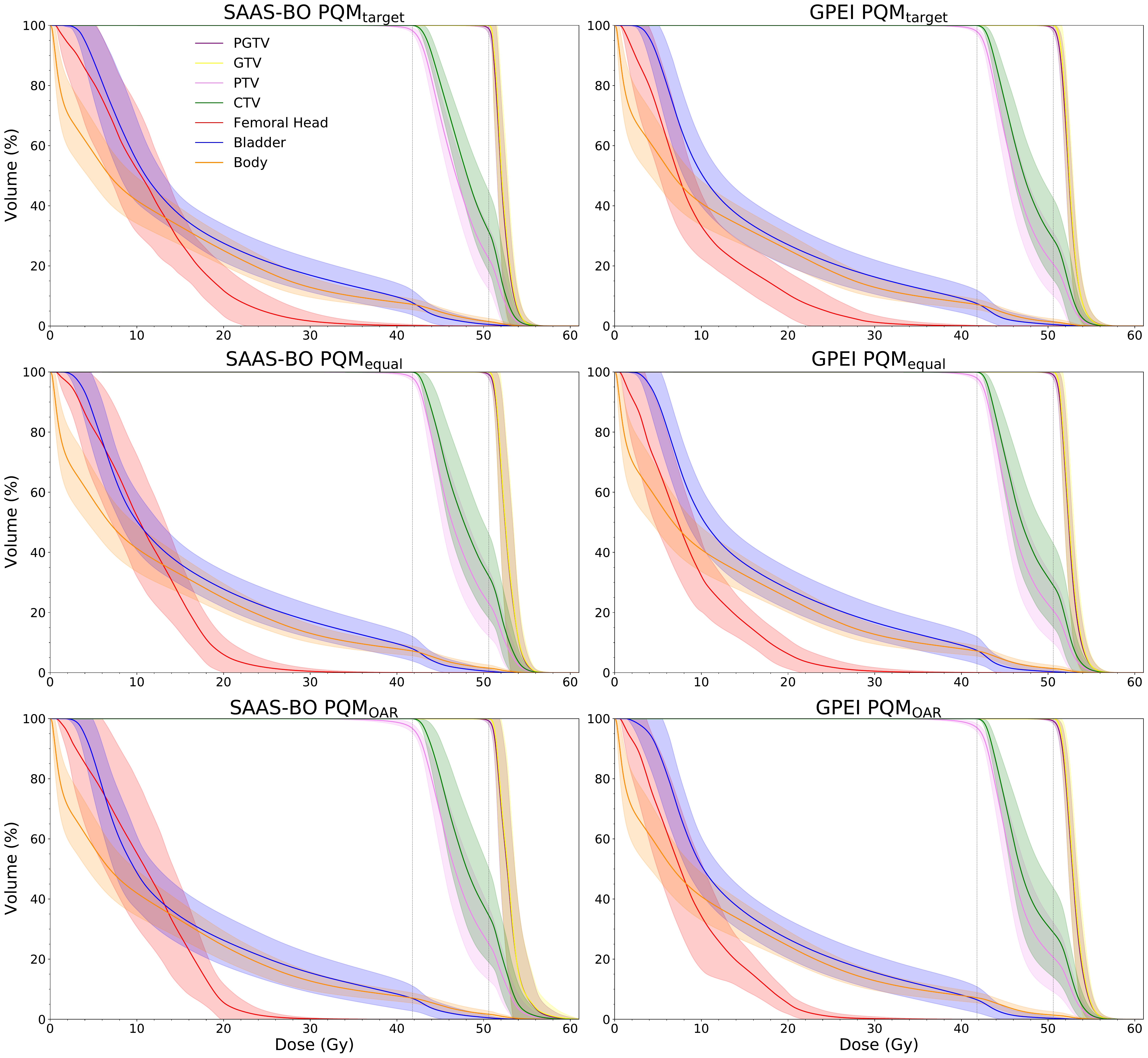}
  \caption{DVH comparison for GPEI and SAAS-BO methods based on different
    clinical priority sets. The solid lines represent the means and the color
    bands represent 1 SD, and dose prescriptions were represented by gray dashed
    lines.}
  \label{fig:8}
\end{figure}

\indent The DVHs with the different sets of priorities were shown in figure \ref{fig:8}. It is easy visible that the control in target hot spot and the maximum dose of the femoral head are susceptible to the influence of changes in the priorities.
 
\begin{table}[H]
\centering
\arrayrulecolor{black}
\caption{Three sets of clinical goal priorities $p_{i}$.}
\label{tab:1}
\begin{tabular}{cccc} 
\arrayrulecolor{black}\hline
 & Target & OAR & Dose Spillage \\ 
\arrayrulecolor{black}\hline
$\mathrm{PQM_{target}}$ & 1 & 2 ($p_{i}$ for femoral head are 3) & 2 \\
$\mathrm{PQM_{equal}}$ & 1 & 1 & 1 \\
$\mathrm{PQM_{OAR}}$ & 2 & 1 & 1 \\
\hline
\end{tabular}
\end{table}

\section{Performance of the Nelder-Mead method given longer planning time}
\begin{figure}[H]
  \centering
  \captionsetup{justification=justified}
  \includegraphics[width=\textwidth]{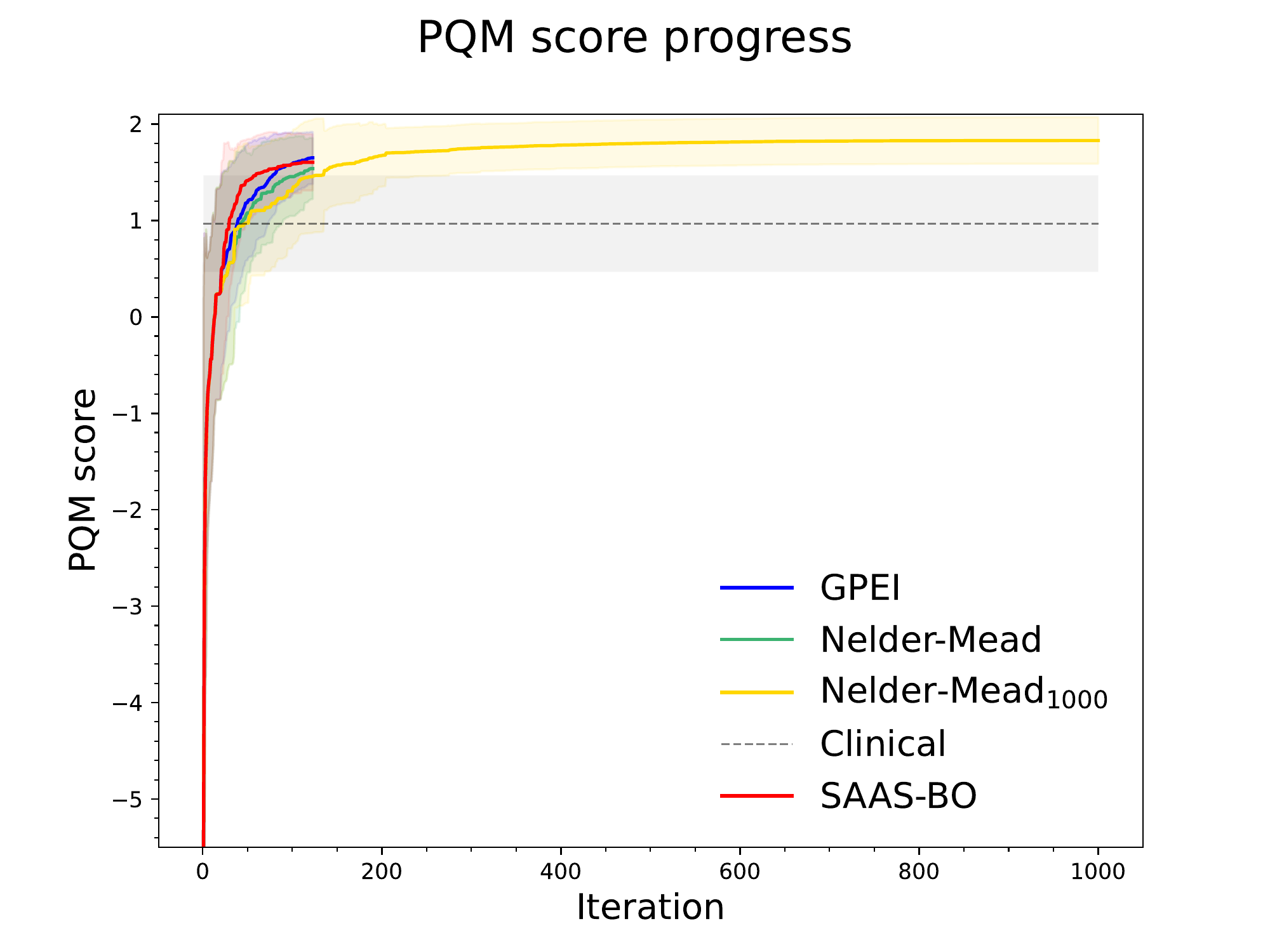}
  \caption{PQM score progress of 1000 iterations of the Nelder-Mead method and 120 iterations of the Nelder-Mead, GPEI and SAAS-BO method along with the average PQM score of clinical treatment plans, where the color bands represent 1 SD.}
  \label{fig:progress_nm1000}
\end{figure}

\begin{figure}[H]
  \centering
  \captionsetup{justification=justified}
  \includegraphics[width=\textwidth]{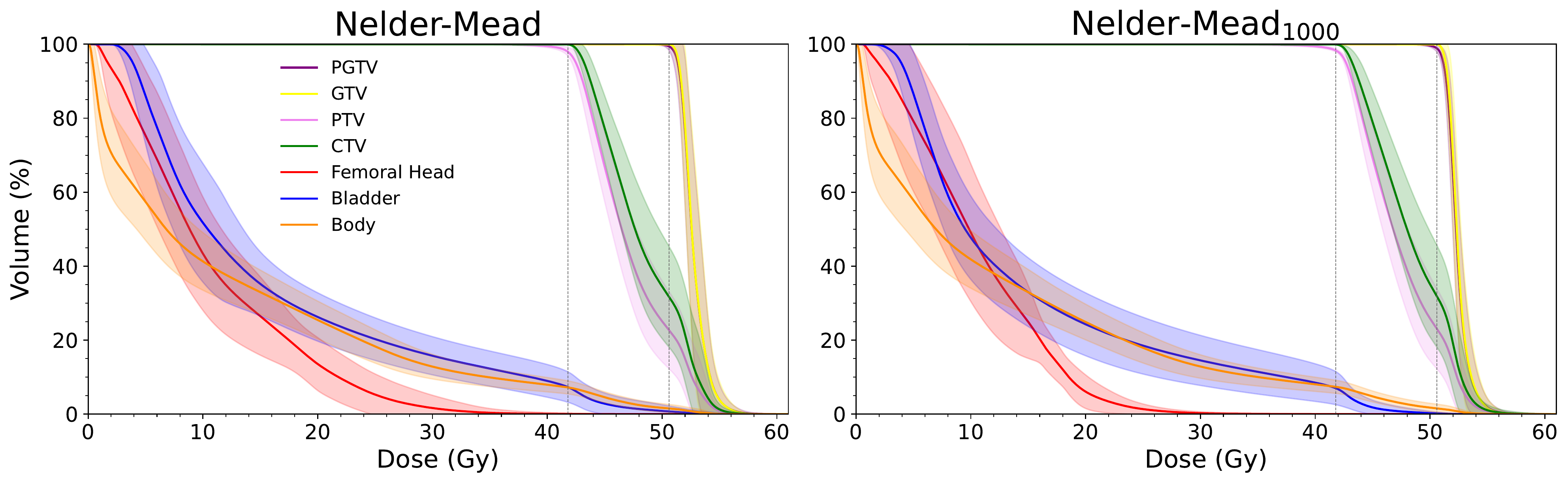}
  \caption{ DVH comparison between 120 iterations of the Nelder-Mead method (right) and 1000 iterations of the Nelder-Mead method (left) on the rectal cancer treatment plan cohort. The solid lines represent the means and the color bands represent 1 SD, and dose prescriptions were represented by gray dashed lines.}
  \label{fig:dvh_nm1000}
\end{figure}

\begin{figure}[H]
  \centering
  \captionsetup{justification=justified}
  \includegraphics[width=\textwidth]{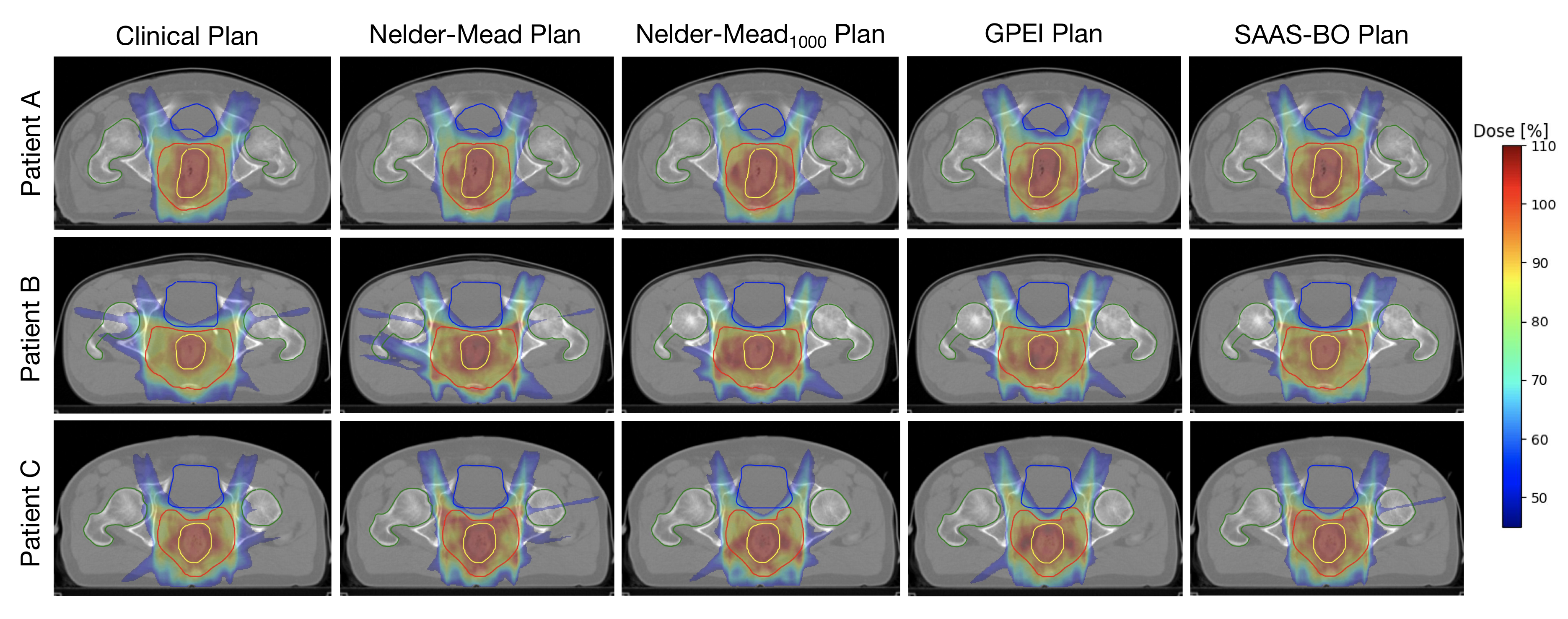}
  \caption{Comparison of dose distributions (100\% to 45\% prescribed dose) for three typical cases with the structures of PGTV(yellow), PTV(red), bladder(blue) and femoral head (green)}
  \label{fig:dosedis_nm1000}
\end{figure}

 \indent In the manuscript, the test methods were limited to 120 iterations, but the planning time of each method varied significantly. Although we explain two factors for the fast runtime of the Nelder-Mead method in the footnote of page 13, we 
have set the iteration limit in the Nelder-Mead method to 1000 (noted as $\rm Nelder-Mead_{1000}$) and repeated the planning experiments on 20 patients to further explore the performance of the Nelder-Mead method. The average best PQM score progress of the Nelder-Mead method compared with previous results was shown in Figure \ref{fig:progress_nm1000} and the mean DVH comparison of the Nelder-Mead method was shown in Figure \ref{fig:dvh_nm1000}. Further, the progress of dose distribution generated by the Nelder-Mead plans also be investigated in the three typical cases (same as the cases in the manuscript) in Figure \ref{fig:dosedis_nm1000}.
\indent As Figure \ref{fig:progress_nm1000} shown, the $\rm Nelder-Mead_{1000}$ plans achieved the same level of average plan quality as the BO plans when iterating to approximately 200 which is twice as many iterations as the BO plans. In subsequent iterations, the average PQM score surpassed that of the BO plans, and the progress was investigated by comparing mean DVH and dose distributions. Figure \ref{fig:dvh_nm1000} clearly demonstrated that the femoral head had the most significant dose reduction and the hot spot control was improved by extending the planning time of the Nelder-Mead method, which were two main contributors to the score progress. However, the Nelder-Mead plans had the problems of hot spots outside the PTV, and more low-dose wash regardless of the number of iterations, which can be visible in 
patient B and C.

